\documentclass[aps,preprint,onecolumn,groupedaddress,nofootinbib,superscriptaddress]{revtex4-1}
\usepackage{graphicx}  
\usepackage{dcolumn}   
\usepackage{bm}        
\usepackage{amsmath,amssymb,amsxtra,bm,epsfig}   
\usepackage{float}
\usepackage[dvipsnames]{xcolor}
\usepackage{xspace}
\usepackage{comment}

  \def \upmns {{\rm{U_{PMNS}}}}

\newcommand{\Dmq}{\Delta m^2}
\newcommand{\dCP}{\delta_\mathrm{CP}}

\newcommand{\dcp}{\delta_{\mathrm{CP}}}
\newcommand{\nova}{NO$\nu$A~}
\newcommand{\tz}{{\theta_{23}}}

\newcommand{\ket}[1]{|{#1}\rangle}
\newcommand{\N}[1]{\widetilde \nu}
\newcommand{\Nm}[1]{\ket{\widetilde \nu^{(m)}_{#1}}}
\newcommand{\Nf}[1]{\ket{\widetilde\nu^{(f)}_{#1}}}
\newcommand{\nuf}[1]{\ket{\nu^{(f)}_{#1}}}
\newcommand{\nuem}[1]{\ket{\nu^{(m)}_{#1}}}

\newcommand\snowmass{\begin{center}\rule[-0.2in]{\hsize}{0.01in}\\\rule{\hsize}{
0.01in}\\
\vskip 0.1in Submitted to the  Proceedings of the US Community Study\\ 
on the Future of Particle Physics {(Snowmass 2021)}\\ 
\rule{\hsize}{0.01in}\\\rule[+0.2in]{\hsize}{0.01in} \end{center}}

\usepackage[linktoc=all, bookmarks, breaklinks, colorlinks=true,urlcolor=RoyalBlue, citecolor=RubineRed, 
linkcolor=ForestGreen]{hyperref}

\usepackage[font=scriptsize]{subcaption}

\begin{document}
	
	\title{Discrete Flavor Symmetries  and Lepton Masses and Mixings}
	\author{Garv Chauhan}
\affiliation{Centre for Cosmology, Particle Physics and Phenomenology (CP3)\\
\centerline{Universit\'{e} catholique de Louvain, 
Louvain-la-Neuve, Belgium}}
\author{P. S. Bhupal Dev}
\affiliation{{Department of Physics and McDonnell Center for the Space Sciences,  Washington University, St. Louis, 
USA
}}
\author{Bartosz Dziewit} 
\affiliation{Institute of Physics, University of Silesia,  Katowice, Poland}
\author{Wojciech Flieger}
\affiliation{
{Max-Planck-Institut f\"ur Physik, Werner-Heisenberg-Institut, 
M\"unchen, Germany
}}
\author{Janusz Gluza} 
	\author{Krzysztof Grzanka} \author{Biswajit Karmakar\footnote{Corresponding author, biswajit.karmakar@us.edu.pl}} \author{Joris Vergeest}
	\author{Szymon Zieba}
	\affiliation{Institute of Physics, University of Silesia,  Katowice, Poland}

	\begin{abstract}
We discuss neutrino mass and mixing models based on discrete flavor symmetries. These models can include a variety of new interactions and non-standard particles such as sterile neutrinos, scalar Higgs singlets and multiplets. We point at connections of the models with leptogenesis and dark matter and the ways to detect the corresponding non-standard particles at intensity and energy frontier experiments.    
	\end{abstract}
{
\let\clearpage\relax
\maketitle
}
\snowmass
\flushbottom
{
\hypersetup{linkcolor=black}
\tableofcontents
}
\newpage

\section{Introduction}
\label{sec:intro}

Over the past few decades we have seen a spectacular progress in understanding neutrino physics.
The neutrino quantum oscillation phenomenon
established that at least two out of three known neutrinos are massive, although their masses are very tiny, at most at the electronvolt level,
$m_{\nu}\sim {\cal{O}}(1)$ eV \cite{Barger:1999na,Czakon:1999cd}. 
It was a tremendous effort that led to this result as experimental studies of neutrino physics face the challenge of low event statistics for an anyhow scarce set of observables. It started around half a century ago with the pioneering Homestake experiment and the so-called solar neutrino problem  \cite{Bahcall:1976zz}, ending with the 2015 year's Nobel award for Takaaki Kajita and Arthur B. McDonald (Super-Kamiokande and SNO Collaborations) \cite{Fukuda:1998mi,Ahmad:2002jz}
 "for the discovery of neutrino oscillations, which shows that neutrinos have mass".
Then and nowadays the neutrino theory is based on the standard parametrization of the mixing matrix through the so-called   \texttt{PMNS} mixing matrix \cite{Pontecorvo:1957qd,Maki:1962mu, Kobayashi:1973fv} 
\begin{align}
  \upmns &=U(\theta_{23})U(\theta_{13})U(\theta_{12})U(\delta_{CP}^M) \nonumber \\
  &=\begin{pmatrix}
    1 & 0 & 0 \\
    0 & c_{23}  & {s_{23}} \\
    0 & -s_{23} & {c_{23}}
  \end{pmatrix}
  \begin{pmatrix}
 c_{13} & 0 & s_{13} e^{-i\delta_\text{CP}} \\
    0 & 1 & 0 \\
    -s_{13} e^{i\delta_\text{CP}} & 0 & c_{13}
  \end{pmatrix}
  \begin{pmatrix}
    c_{12} & s_{12} & 0 \\
    -s_{12} & c_{12} & 0 \\
    0 & 0 & 1
  \end{pmatrix}  \begin{pmatrix}
    e^{i \alpha_1} & 0 & 0 \\
    0 & e^{i \alpha_2} & 0 \\
    0 & 0 & 1
  \end{pmatrix},  
\label{upmns1}
\end{align}

\!\!where we denote $c_{ij} \equiv \cos( \theta_{ij})$, $s_{ij} \equiv \sin(\theta_{ij})$,
and the Euler rotation angles $\theta_{ij}$ can be taken without loss
of generality from the first quadrant, $\theta_{ij} \in [0, \pi/2]$, and the standard $CP$ phase $\delta_{CP} \in [0, 2\pi]$. This choice of parameter regions is independent of matter effects \cite{Gluza:2001de}. CP-phases $\alpha_1$ and $\alpha_2$ are present for Majorana neutrinos.

Tab.~\ref{neutrino_data} shows recent global fits for the neutrino parameters, which are used in present analyses for two mass orderings, defined as:
\begin{equation}
\begin{array}{ccc}
\mbox{Normal mass hierarchy (NH):} & \mbox{Inverted mass hierarchy (IH):} \\
& \\
\begin{array}{l}
m_{\nu_1} = m_{0}, \\
m_{\nu_2} = \sqrt{m_{0}^2 + \Delta m_{21}^2}, \\
m_{\nu_3} = \sqrt{m_{0}^2 + \Delta m_{31}^2}, \\
\end{array}
&
\begin{array}{l}
m_{\nu_1} = \sqrt{m_{0}^2 - \Delta m_{21}^2 - \Delta m_{32}^2}, \\
m_{\nu_2} = \sqrt{m_{0}^2 - \Delta m_{32}^2}, \\
m_{\nu_3} = m_{0}, 
\end{array}
\end{array}
\label{e:NH_and_IH}
\end{equation}
where $\Delta m_{ij}^2 \equiv m_i^2-m_j^2$.
In inverse hierarchy/ordering (IH/IO), the mass of $\nu_3$ is smaller than masses of the $\nu_1,\nu_2$ states. The solar oscillation effect is between $\nu_1,\nu_2$ states for both hierarchies, so in the spectrum with normal hierarchy/ordering (NH/NO) $m_1 < m_2 < m_3$, while in 
IH $m_3 < m_1 < m_2$. For both orders $m_0$ indicates the smallest of the three neutrino masses.
Furthermore, the data show a hierarchy between the mass splittings, $\Delta m_{21}^2 \ll \Delta m_{31}^2 \simeq \Delta m_{32}^2$. 

\begin{table}[H]
\begin{center}
    \begin{tabular}{c}
         \includegraphics[scale=0.65]{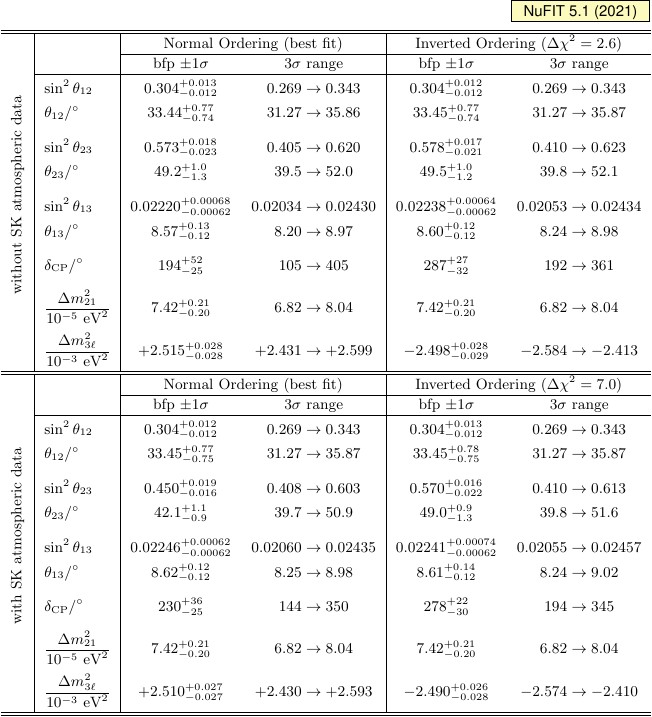}
    \end{tabular}
\end{center}
\caption{Neutrino oscillation data, $\Delta m_{ij}^2 \equiv m_i^2-m_j^2$. Depending on the hierarchy, either $\Delta m_{3l}^2 \equiv \Delta m_{31}^2 > 0$ (NH) or \mbox{$\Delta m_{3l}^2 \equiv \Delta m_{32}^2 < 0$~(IH)} \cite{NuFIT5.1,Esteban:2020cvm}.
}
 
\label{neutrino_data}
\end{table}


The T2K data exclude CP conservation in neutrino oscillations at a $2\sigma$ level.
The initial important results by the T2K collaboration \cite{Abe:2019vii} 
show an indication of CP violation in the lepton sector, 
a preference for  NH at the 3$\sigma$ level, and also for $\theta_{23}$ belonging to the second octant. 
These data are confirmed with an improved analysis \cite{T2K:2021xwb}; they show a weak preference for normal mass ordering (89\% posterior probability) and the upper  octant (80\% posterior probability), with a uniform prior probability assumed in both cases.
On the other hand, the moderate excess at \nova \cite{NOvA:2021nfi} leads to two degenerate solutions either for NH with $0<\dcp<180^\circ$ and $\tz>\pi/4$, or  IH with $-180^\circ<\dcp<0$ and $\tz>\pi/4$. The \nova data disfavor combinations of oscillation parameters that give rise to a large asymmetry in the appearance rates of $\nu_e$ and $\bar{\nu}_e$. 
This implies values of the CP-violating phase in the vicinity of $\delta_{CP}=\pi/2$, disfavored by 3$\sigma$ for IH, and values around $\delta_{CP}=3\pi/2$ in NH, disfavored at 2$\sigma$ for IH.
For a recent review, see \cite{Rahaman:2022rfp}.
Let us note that the T2K, Super-Kamiokande and the \nova collaborations are working towards joint analyses. \cite{T2K:2021xwb}.
Oscillation experiments do not put a limit on the Majorana phases $\alpha_1$, $\alpha_2$. 
However, in some studies there are predictions for the Majorana phases using the neutrinoless double beta decay, e.g., see \cite{Girardi:2016zwz} and Eq.~(\ref{2nu0}) below.  
There is no bound on the individual masses of neutrinos from the oscillation data. Therefore, the lightest neutrino mass $m_{0}$ is a free parameter and the other two masses are determined through (\ref{e:NH_and_IH}). Also, there are limits on the sum of three neutrino masses from different experiments: from tritium decay \cite{Zyla:2020zbs} or neutrinoless double beta decay~\cite{Rodejohann:2011mu}, and astrophysics and cosmology.

In tritium beta decay, an effective parameter is $m^2_{\nu_e}
    = \frac{\sum_i m^2_i |U_{ei}|^2}{\sum_i |U_{ei}|^2}=\sum_i m^2_i |U_{ei}|^2$ which can be expressed through oscillation parameters \cite{Esteban:2018azc}

\begin{equation}
  \label{eq:mbeta}
  \begin{aligned}
    m^2_{\nu_e}
    &=  
     c_{13}^2 c_{12}^2 m_1^2
    + c_{13}^2 s_{12}^2 m_2^2+s_{13}^2 m_3^2
    = \begin{cases}
      \text{NH: }
      & m^2_0 + \Dmq_{21} c_{13}^2 s_{12}^2+\Dmq_{3\ell} s_{13}^2 \,,
      \\
      \text{IH: }
      & m^2_0 - \Dmq_{21} c_{13}^2 c_{12}^2-\Dmq_{3\ell} c_{13}^2.
    \end{cases}
  \end{aligned}
\end{equation}

At present only an upper bound is known, $m_{\nu_e} \leq 2.2$~eV at 95\%
CL~\cite{Bonn:2001tw}, which has been superseded recently by
KATRIN~\cite{KATRIN:2001ttj} with an upper bound, $m_{\nu_e} < 0.8$~eV at 90\%
CL~\cite{Aker:2022}. 

In neutrinoless double beta decay, an effective parameter is $m_{ee} = \Big| \sum_i m_i U_{ei}^2 \Big|$ which can be expressed through oscillation parameters \cite{Esteban:2018azc}

\begin{multline}
  m_{ee} 
  = \Big| m_1 c_{13}^2 c_{12}^2 e^{i 2\alpha_1} +
  m_2 c_{13}^2 s_{12}^2 e^{i 2\alpha_2} +
  m_3 s_{13}^2 e^{-i 2\dCP} \Big|
  \\
  = \begin{cases}
    \text{NH:}
    & m_0\, \Big|c_{13}^2 c_{12}^2 e^{i 2(\alpha_1-\dCP)}
    + \sqrt{1 + \frac{\Dmq_{21}}{m_0^2}} \, c_{13}^2 s_{12}^2 e^{i 2(\alpha_2-\dCP)}
    + \sqrt{1 + \frac{\Dmq_{3\ell}}{m_0^2}}\, s_{13}^2 \Big|
    \\
    \text{IH:}
    & m_0\, \Big| \sqrt{1-\frac{\Dmq_{3\ell} +\Dmq_{21}}{m_0^2}} \,
    c_{13}^2 c_{12}^2 e^{i 2(\alpha_1 - \dCP)}
    + \sqrt{1-\frac{\Dmq_{3\ell}}{m_0^2}}\,
    c_{13}^2 s_{12}^2 e^{i 2(\alpha_2-\dCP)} + s_{13}^2 \Big|.
  \end{cases} \label{2nu0}
\end{multline}

The present limit on the neutrino mass of $m_{ee}$ is $\lesssim 0.06-0.200$ eV at 90\% CL. \cite{KamLAND-Zen:2016pfg,Dolinski:2019nrj}.  
New experiments nEXO~\cite{nEXO:2021ujk} and LEGEND~\cite{LEGEND:2021bnm}    are planned, with sensitivity up to $m_{ee} \sim
0.01$~eV~\cite{wwwubound}, more specifically: $m_{ee} < \left(4.7 -20.3\right)$\;{meV}  {{(nEXO)}},
        $m_{ee} < \left(34-78\right)$\;{meV}  { {(LEGEND-200)}} and
        $m_{ee} < \left(9-21\right)$\;{meV}  {\textrm {(LEGEND-1000)}}. 

{Finally, cosmological data gives at 95\% CL \cite{Planck:2018nkj,eBOSS:2020yzd}}
\begin{equation}
\Sigma \equiv \sum_{i=i}^3 m_{\nu_i} < 0.12 \; \mbox{eV}.
\label{astrolimit}
\end{equation}

These limits set the upper bound on the lightest neutrino mass \cite{Barger:1999na,Czakon:2001uh,Pascoli:2002xq}.  
Further, taking into account data in Tab.~\ref{neutrino_data} and (\ref{astrolimit}), we get even more strict upper limits on $m_0$ for NH and IH, at the level of 50 meV.

  
Now understanding the pattern of neutrino mixing is very crucial because it is essentially part of the long-standing flavor puzzle and intrinsically different from
that of quark mixing.  Particularly, the  patterns of neutrino mixing and origin of their masses remain elusive  and {go beyond the Standard Model theory}. 
As can be seen in Tab.~\ref{neutrino_data}, the global analysis of neutrino oscillation data suggests that the two mixing angles ($\theta_{23}$ and $\theta_{12}$) are relatively large compared to the reactor mixing angle $\theta_{13}$. In fact prior to 2012 for a long period of time,  the reactor mixing angle $\theta_{13}$ was thought to be vanishingly small. Over the time guided by $\theta_{13} \sim 0^{\circ}$ and to be consistent with the observed data, several mixing schemes were postulated to explain the observed pattern of the lepton mixing matrix. By substituting $\theta_{13}=0^{\circ}$ and $\theta_{23}=45^{\circ}$ in the general lepton mixing matrix given in Eq.~(\ref{upmns1}), most of the popular mixing schemes such as bi-maximal mixing (BM)~\cite{Vissani:1997pa, Barger:1998ta, Baltz:1998ey}, tribimaximal mixing (TBM)~\cite{Harrison:2002er, Harrison:2002kp}, golden ratio (GR)~\cite{Datta:2003qg, Kajiyama:2007gx, Everett:2008et, Rodejohann:2008ir, Adulpravitchai:2009bg, Ding:2011cm} and hexagonal (HG)~\cite{Albright:2010ap} mixings can be altogether written as\footnote{There will an additional phase matrix $U_p$ with this matrix if neutrinos are Majorana particles.} 
\begin{eqnarray}
				U_0=	\left(
					\begin{array}{ccc}
						c_{12} &s_{12} &0\\
						-\frac{s_{12}}{\sqrt{2}} &\frac{c_{12}}{\sqrt{2}} &-\frac{1}{\sqrt{2}}\\
						-\frac{s_{12}}{\sqrt{2}} &\frac{c_{12}}{\sqrt{2}} &\frac{1}{\sqrt{2}}
					\end{array}
					\right).\label{ugen}
				\end{eqnarray}
Here BM, TBM, GR\footnote{There exists an alternate version of GR mixing where $\cos\theta_{12}=\phi/2$ \cite{Rodejohann:2008ir, Adulpravitchai:2009bg}.}, HG mixing schemes can be obtained by substituting $s_{12}=1/\sqrt{2}$, $s_{12}=1/\sqrt{3}$, $\tan\theta_{12}=1/\phi$ where $\phi=(1+\sqrt{5})/2$ and $s_{12}=1/2$, respectively. Using the diagonalization relation 
\begin{align}
    m_\nu=U_0^\star {\rm diag} (m_1,m_2,m_3)U_0^{\dagger}, 
\end{align}
such a mixing matrix  can  easily diagonalize a $\mu-\tau$ symmetric (transformations $\nu_{e} \rightarrow\nu_{e}$, $\nu_{\mu}\rightarrow\nu_{\tau}$, $\nu_{\tau}\rightarrow\nu_{\mu}$ under which the neutrino mass term remains unchanged) neutrino mass matrix of the form  \begin{eqnarray}
			m_\nu=	\left(
					\begin{array}{ccc}
						A &B &B\\
					B & C & D\\
						B &D  &C
					\end{array}\label{mmutau}
					\right), 
\end{eqnarray}
where the elements $A,B,C$ and $D$ in Eq. (\ref{mmutau}) are in general complex. With $A+B=C+D$ this matrix yields TBM mixing pattern where $s_{12}=1/\sqrt{3}$~\cite{Harrison:2002er, Harrison:2002kp}.  Such first order approximations of the neutrino oscillation data stimulated  theorists to find an aesthetic  framework which can  lead  towards these fixed mixing matrices. In this regard non-Abelian discrete groups\footnote{Discrete groups have always played a key role in physics starting from crystallographic groups in solid state physics, to discrete symmetries such as $C$, $P$, and $T$, and their violation have shaped our understanding of nature.} turned out to be popular as appropriate flavor symmetries for the lepton sector. For a long time, various discrete groups such as $S_3$, $A_4$, $S_4$, $A_5$, $T'$, $\Delta(27)$, $D_n$, $T_{7}$, $\Delta(6n^2)$.  have been extensively used  to explain fermionic mixing. The most popular discrete group was initially proposed as an underlying family symmetry for the quark sector \cite{Wyler:1979fe, Branco:1980ax}. For a general introduction see~\cite{Frampton:1994rk} and for implementation of discrete symmetries  in neutrino physics the readers are refereed to ~\cite{Ishimori:2010au, Altarelli:2010gt, Feruglio:2019ybq, King:2013eh, Petcov:2017ggy, Xing:2020ijf} and the references therein.  Interestingly,  in the last decade, thanks to the various  neutrino experiments like  Double Chooz~\cite{Abe:2011fz}, Daya Bay~\cite{An:2012eh}, RENO~\cite{Ahn:2012nd}, T2K~\cite{T2K:2013ppw}, MINOS~\cite{MINOS:2013utc},  {and others \cite{wwwubound}}, the reactor neutrino mixing angle is conclusively  measured to be  substantially large ($\theta_{13}\sim 8^{\circ}-9^{\circ}$). In addition to this, as mentioned earlier, a non-zero value of the Dirac CP phase  $\delta_{CP}$ is favoured by the oscillation experiments. Such {an} observation has ruled out  the possibility {of} having neutrino mixing schemes like BM, TBM, GR, HG.  Therefore, modification or corrections or figuring out the successors  of the above mixing schemes (which are still viable) is consequential and discussed in the next section. Models based on non-Abelian discrete flavor symmetries often yield interesting predictions and correlations among the neutrino masses, mixing angles and CP phases. 
~Involvement of such studies   may have wider applications in various aspects of cosmology (matter-antimatter asymmetry of the Universe, dark matter, gravitational waves etc.), collider physics and other aspects in particle physics.  

Moreover, as the light neutrino sector and masses connected with three families of neutrinos and charged leptons can be a reflection of the more general theory where weakly interacting (sterile) neutrinos exist with much higher masses (GeV, TeV, up to the GUT scale and beyond), a related problem is the symmetry of the whole neutrino mass matrix or its heavy sector part. 

In this framework, the established neutrino mass and flavor states can be denoted by ${ \nuem \alpha }$ and 
${ \nuf \alpha }$, respectively. Any extra, beyond SM (BSM) mass and flavor states we denote by $\Nm j$ and $\Nf j$ for $j=1,\ldots, n_R$, respectively.
In this general scenario mixing between an extended set of neutrino mass states $\{\nuem \alpha,\Nm \beta\}$ with flavor states $\{\nuf \alpha,\Nf \beta\}$ is described by
\begin{align}
\begin{pmatrix} { \nuf \alpha} \\ 
 \Nf \beta\end{pmatrix}  &=
 \begin{pmatrix} {{ \upmns }} & V_{lh} \\  V_{hl} & V_{hh}
  \end{pmatrix} 
\begin{pmatrix} {\nuem \alpha } \\ 
\Nm \beta\end{pmatrix} 
\equiv U  \begin{pmatrix} {{ \nuem \alpha }} \\ \Nm \beta\end{pmatrix}\;.
\label{ugen}
\end{align}

The observable part of the above is the transformation from mass $\nuem \alpha, \Nm \beta$ to SM flavor $\nuf \alpha$ states 
\begin{equation}
\nuf \alpha = 
\sum_{i=1}^3\underbrace{\left( \upmns  \right)_{\alpha i} {\nuem i}}_{\rm SM \;part}
+ 
\sum_{j=1}^{n_R}\underbrace{
{ \left(V_{lh}  \right)}_{\alpha j} \Nm j}
_{\rm BSM \; part}\;. \label{gen}
\end{equation}  

The mixing matrix $U$ in (\ref{ugen}) diagonalizes general neutrino mass matrix 
\begin{equation}\label{ssmm}
M_{\nu}=
\left(
\begin{array}{cc}
M_L & M_{D} \\
M_{D}^T & M_{R} 
\end{array}
\right) ,
\end{equation}
\noindent
using a congruence transformation
\begin{equation}
U^{T}M_{\nu}U \simeq diag(M_{light}, M_{heavy}).
\label{congr}
\end{equation}

Certainly the structure and symmetry of the heavy neutrino sector $M_R$ embedded in $M_\nu$ in Eq.~(\ref{ssmm}), altogether with  $M_\nu$ variants and extensions, influence the masses and mixings of the light sector, in the first row in the seesaw type of models \cite{Flieger:2020lbg}.

Now we will discuss different ways of linking various flavor symmetries with neutrino mass and mixing matrices.  
  
\section{Flavor Symmetry  and Lepton Masses and Mixing}

From Eq.~(\ref{upmns1}) we find that the neutrino mixing matrix is expressed in terms of  few mixing angles and CP violating phases and we are yet to understand the experimentally observed mixing pattern~\cite{Esteban:2020cvm}. The masses and mixing of the  leptons (as well as of the quarks) are obtained from the Yukawa couplings related to the families. Therefore, {\it{it is natural to ask whether such a mixing pattern is  governed by any fundamental principle.}}  
\subsection{General Framework}
The primary approaches which try to address the issue of the neutrino mixing pattern include (i) random analysis without imposing prior theories or symmetries on the mass and mixing matrices \cite{Hall:1999sn,deGouvea:2012ac,Haba:2000be}; (ii) more specific studies with imposed mass or mixing textures for which models with underlying symmetries can be sought \cite{Xing:2002ta,Zhou:2004wz,Dziewit:2006cg,Ludl:2014axa,Fukugita:2016hzf}, and finally, (iii) theoretical studies where some explicit symmetries at the Yukawa Lagrangian level are assumed and corresponding extended particle sector is defined. 
In the anarchy hypothesis (i) 
leptonic mixing matrix is a manifestation of a random draw from an unbiased distribution of unitary  $3 \times 3$ matrices and do not point towards any principle or its origin. This hypothesis does not make any correlation among the neutrino masses and mixing parameters. However, it  predicts probability distribution for the  parameters which parameterizes the mixing matrix. 
{Though random matrices cannot solve
fundamental problems in neutrino physics, they generate
intriguing hints on the nature of neutrino mass matrices. For instance, in \cite{Gluza:2011nm} preference has been observed towards random models of neutrino masses with sterile neutrinos}.
In the intermediate approach (ii) some texture zeros of neutrino mass matrices can be eliminated. 
For instance, in \cite{Ludl:2014axa} 570 (298)
inequivalent classes of texture zeros in the Dirac (Majorana) case were found.  
For both cases about 75\% of the classes are compatible with the data. In case of maximal texture zeros in the neutrino and charged lepton mass matrices there are only 
about 30 classes of texture zeros for each of the four categories defined by Dirac/Majorana
nature and normal/inverted ordering of the neutrino mass spectrum. 
A strict texture neutrino mass matrix can be also discussed in the context of phenomenological studies, for instance leptogenesis \cite{Fukugita:2016hzf}.
  
{In what follows, we will discuss} the symmetry based approach (iii) to explain the non-trivial mixing in the lepton sector known as family symmetry or horizontal symmetry. Such fundamental symmetry in the lepton sector can easily explain the origin of neutrino mixing which {is} considerably different from quark mixing.  Incidentally, both  Abelian or non-Abelian family symmetries have potential to shade light on the Yukawa couplings. The Abelian symmetries (such as Froggatt-Nielsen symmetry \cite{Froggatt:1978nt}) only points towards a hierarchical structure of the Yukawa couplings whereas non-Abelian {symmetries} are more equipped to explain the   non-hierarchical structures of the observed lepton mixing as observed by the oscillation experiments. 

If we consider a family symmetry $G_f$, the three generations of leptons and quarks can be assigned to irreducible representations or multiplets, hence unifying the flavor of the generations. If $G_f$ contains a triplet representation $\mathbf{3}$, all three fermion families  can follow {the} same transformation properties. For example, let us consider that non-zero neutrino mass is generated through the Weinberg operator $H L^T c LH $ where the lepton and Higgs doublets transform as triplet ($\mathbf{\bar{3}}$) and singlet under a family symmetry, say,   $SU(3)$.   Now to construct a $SU(3)$ invariant operator, an additional scalar fields $\Phi$  (also known as {\it{flavon}} ) is  introduced and the effective operator takes the form $H L^T \Phi^T \Phi LH $. {A} suitable vacuum  alignment ($\langle \Phi \rangle \propto (u_1, u_2, u_3)^T$) for the flavon   is inserted in such a way that the obtained mass matrix is capable of appropriate mixing pattern.
As a result $G_f$ is spontaneously broken once  these flavons acquire non-zero vacuum expectation values (VEV). Continuous family symmetry such as $U(3)$, $O(3)$ (and their subgroups $SU(3)$ and $SO(3)$) can in principle be used for this purpose
to understand the neutrino mixing. 
However non-Abelian discrete flavor symmetric approach is much more convenient as in such framework obtaining  the desired vacuum alignment (which produces correct mixing) of the flavon can be obtained easily~\cite{Altarelli:2005yp, Altarelli:2005yx}. 
At this point it is worth mentioning that, these  non-Abelian discrete symmetries can also originate from a continuous symmetry~\cite{Koide:2007sr,Merle:2011vy, Adulpravitchai:2009kd, Luhn:2011ip, Wu:2012ria, King:2018fke, Ludl:2010bj, Joshipura:2016quv, Joshipura:2015dsa}. For example widely used discrete groups such as $A_4, S_4, A_5, \Delta(27), T_7$  can originate from the  {continuous} group $SU(3)$ \cite{King:2013eh}. In another example~\cite{King:2018fke} the authors showed that continuous $SO(3)$  can also give rise to $A_4$ which is further broken to smaller $Z_3$ and $Z_2$ symmetries.  Few years back it was proposed that various non-Abelian discrete symmetries can also originate   from superstring theory through compactification of extra dimensions and known as {the} modular invariance approach~\cite{Altarelli:2005yx, deAdelhartToorop:2011re, Feruglio:2017spp}.

In this  report we will concentrate on all these aspects of discrete family symmetries discussed above and their implications to understand the lepton mixing and its extensions.
The model building with flavor symmetries is not  trivial  since the underlying flavor symmetry group $G_f$
must be broken. Usually this symmetry $G_f$ is considered to exist at some high  scale (sometimes with a close proximity to GUT scale~\cite{Altarelli:2010gt}) and to be broken at lower energies
with residual symmetries of the charged lepton and neutrino sectors, represented by the subgroups $G_e$ and $G_\nu$, respectively. Therefore, the choice of the non-Abelian discrete group $G_f$ and its breaking pattern to yield  remnant   subgroups $G_e$ and $G_\nu$ shapes the model building significantly to obtain definite predictions and correlations of the mixings. In the absence of any residual symmetry, the flavor $G_f$ loses its predictivity markedly. For a detailed discussion on the choice of various discrete symmetries and their generic predictions, see~\cite{King:2013eh, Petcov:2017ggy, Tanimoto:2015nfa}. 
\begin{table}[]
    \centering
    \begin{tabular}{|c |c |c |c| c|}
			\hline
			 Group & Order  & Irreducible Representations  & Generators \\
			\hline
			$A_4$ & 12  & $1, 1', 1'', 3$  &  $S,T$ \\ 
			$S_4$ &  24 & $1,1',2,3,3'$  &   $S,T(U)$ \\
			$T'$ & 24 &  $1,1',1'',2,2',2'',3$      & $S,T(R)$  \\
			$\Delta(27)$ & 27 & $1_{r,s} (r,s=0,1,2), 3_{01, 02}$ & $C,D$ \\
			$A_5$ & 60 & $1,3,3',4,5$& $\Tilde{S}$, $\Tilde{T}$\\
			\hline
    \end{tabular}
    \caption{Details of {a} few small groups with triplet irreducible representations~\cite{Ishimori:2010au}}
    \label{tab:sg}
\end{table}
In Tab.~\ref{tab:sg} we mention the basics details such as order or number of elements (first column), irreducible representations (second column) and  generators (third column) of  small groups (which contains at least one triplet) such as $A_4,  S_4, T'$  $\Delta(27)$ and $ A_5$. A pedagogical review including catalogues of the generators and multiplication rules  of  these widely  used non-Abelian discrete groups  can be found in~\cite{Ishimori:2010au}.

Now for model building {purposes}, there {exist} various approaches based on the breaking pattern of $G_f$ into its residual symmetries, also known as  direct, semi-direct and indirect approaches ~\cite{King:2013eh, Tanimoto:2015nfa}. After breaking of $G_f$,  different residual symmetries {exist} for charged lepton (typically $G_{e}=Z_3$) and neutrino sector (typically  $G_{\nu}=Z_2 \times Z_2$, also known as the Kline symmetry). It is known as {the} direct approach. In {a} semi-direct approach one of the generators of the residual symmetry is assumed to be broken. {On the} contrary, in the indirect approach, no residual  symmetry of flavor groups remains intact and  the flavons acquire special vacuum alignments whose alignment is guided by the flavor symmetry. Usually, different flavons take part in  the charged lepton and neutrino {sectors}. To show how the family symmetry {shapes} the flavor model building let us consider $G_f=S_4$ as {a} guiding symmetry. Geometrically, this group can be seen as the symmetry group of a rigid cube and it is a group of permutation 4 objects. Therefore, the order of the group is $4!=24$ and  the  elements can be conveniently generated  by the generators $S,T$ and $U$ satisfying the relation 
\begin{equation}
    S^2=T^3=U^2=1~~{\rm and}~~ST^3=(SU)^2=(TU)^2=1. 
\end{equation}
In its  irreducible triplet representations  these three generators  can be written as 
\begin{eqnarray}
				S=	\frac{1}{3}\left(
					\begin{array}{ccc}
						-1 &2 &2\\
						2 &-1 &2\\
						2 &2 &-1
					\end{array}
					\right);   
					T=	\left(
					\begin{array}{ccc}
						1 &0 &0\\
						0 &\omega^2 &0\\
						0 &0 &\omega
					\end{array}
					\right) ~~{\rm and }~~
					U=	\mp \left(
					\begin{array}{ccc}
						1 &0 &0\\
						0 & 0 &1\\
						0 &1 & 0
					\end{array}
					\right). \label{STU}
				\end{eqnarray}
In the direct approach the charged lepton mass matrix ($M_\ell$) respects the generator $T$ whereas the neutrino mass matrix ($M_{\nu}$)  respects the generators $S,U$ satisfying the condition  
\begin{equation}
    T^{\dagger}M_{\ell}^{\dagger}M_{\ell}T=M_{\ell}^{\dagger}M_{\ell},~ S^T M_{\nu} S = M_{\nu}~{\rm and }~U^T M_{\nu} U = M_{\nu}, \label{eq:s4tbm1}
\end{equation}
which leads to~\cite{Tanimoto:2015nfa}
\begin{equation}
    [T, M_{\ell}^{\dagger}M_{\ell}]=[S,M_{\nu}]=[U, M_{\nu}]=0\label{eq:s4tbm2}. 
\end{equation}
Clearly, the non-diagonal matrices $S, U$ can be  diagonalized by a mixing matrix of the form given in Eq. (\ref{ugen}) and can be written as 
\begin{eqnarray}
				U_{\rm TBM}=	\left(
					\begin{array}{ccc}
						\frac{2}{\sqrt{6}} &\frac{1}{\sqrt{3}} &0\\
						-\frac{1}{\sqrt{6}} &\frac{1}{\sqrt{3}} &-\frac{1}{\sqrt{2}}\\
						-\frac{1}{\sqrt{6}} &\frac{1}{\sqrt{3}} &\frac{1}{\sqrt{2}}
					\end{array}
					\right),\label{utbm}
				\end{eqnarray}
which is basically the TBM mixing matrix with  $s_{12}=1/\sqrt{3}$. For generic features of semi-direct and indirect approaches to the  flavor model building we refer the readers to~\cite{King:2013eh, King:2020ldi, Tanimoto:2015nfa}.   The TBM mixing pattern explained here  can be generated using various discrete groups, for detailed models and groups see $A_4$~\cite{Ma:2001dn,Ma:2004zv, Altarelli:2005yp,Altarelli:2005yx}, $S_4$~\cite{Bazzocchi:2012st, Bazzocchi:2008ej}, $\Delta(27)$~\cite{deMedeirosVarzielas:2006fc}, $T'$~\cite{Frampton:2008bz} . In addition,  explicit  models with discrete flavor symmetry for BM~\cite{Girardi:2015rwa,Lam:2008rs,Frampton:2004ud, Merlo:2011vc}, GR~\cite{Everett:2008et, Feruglio:2011qq, Ding:2011cm}, HG~\cite{Kim:2010zub,} mixing can easily be constructed.   

\subsection{Family Symmetry,  nonzero $\theta_{13}$ and nonzero $\delta_{CP}$}
After precise measurement of non-zero value of the reactor mixing angle $\theta_{13}$~\cite{Abe:2011fz, An:2012eh, Ahn:2012nd, MINOS:2013utc, T2K:2013ppw} the era of fixed patterns (such as BM, TBM, GR, HG mixing) of the lepton mixing matrix is over. Also, as mentioned earlier,  long baseline neutrino oscillation experiments such as T2K~\cite{T2K:2019bcf} and NO$\nu$A \cite{NOvA:2019cyt} both hints for CP violation in the lepton sector. Therefore, each of the fixed patterns need some modification to be consistent with global fit of the neutrino oscillation data~\cite{Esteban:2020cvm}. There are many  possible ways to achieve this. For example,  even if the TBM mixing is obsolete now, there are two of its successors which are still compatible with data. These are called TM$_1$ and TM$_2$ mixing and {are} given by 
\begin{eqnarray}
	|U_{{\rm TM}_1}|=	\left(
					\begin{array}{ccc}
						\frac{2}{\sqrt{6}} &* &*\\
						\frac{1}{\sqrt{6}} &* &*\\
						\frac{1}{\sqrt{6}} &* &*
					\end{array}
					\right)~{\rm and}~|U_{{\rm TM}_2}|=	\left(
					\begin{array}{ccc}
						* &\frac{1}{\sqrt{3}} &*\\
						* &\frac{1}{\sqrt{3}} &*\\
						* &\frac{1}{\sqrt{3}} &*
					\end{array}
					\right), \label{utm}
				\end{eqnarray}
respectively. Clearly,  Eq. (\ref{utm}) shows that TM$_1$ and TM$_2$ mixings preserve first and second column of the TBM mixing matrix given in Eq. (\ref{utbm}). Here the reactor mixing angle becomes a free parameter and the solar mixing angle can stick close to its TBM prediction. To illustrate this, let us again consider the discrete flavor symmetry $G_f=S_4$. In contrast to the breaking pattern mentioned in Eqs.~(\ref{eq:s4tbm1}, \ref{eq:s4tbm2}), if  $S_4$ is considered to be broken spontaneously into $Z_3=\{1, T, T^2\}$ (for the charged lepton sector) $Z_{2}=\{ 1, SU\}$ (for the neutrino sector) such that it satisfies  
\begin{equation}
    [T, M_{\ell}^{\dagger}M_{\ell}]=[SU,M_{\nu}]=0\label{eq:s4tbm3}, 
\end{equation}
following the above prescription, the effective mixing matrix can be written as 
\begin{eqnarray}
				U_{{\rm TM}_1}=	\left(
					\begin{array}{ccc}
						\frac{2}{\sqrt{6}} &\frac{c_{\theta}}{\sqrt{3}} &\frac{s_{\theta}}{\sqrt{3}}e^{-i\gamma}\\
						-\frac{1}{\sqrt{6}} &\frac{c_{\theta}}{\sqrt{3}}-\frac{s_{\theta}}{\sqrt{2}}e^{i\gamma} &-\frac{s_{\theta}}{\sqrt{3}}e^{-i\gamma}-\frac{c_{\theta}}{\sqrt{2}}\\
						-\frac{1}{\sqrt{6}} &\frac{c_{\theta}}{\sqrt{3}}-\frac{s}{\sqrt{2}}e^{i\gamma} &-\frac{s_{\theta}}{\sqrt{3}}e^{-i\gamma}+\frac{c_{\theta}}{\sqrt{2}}
					\end{array}
					\right),\label{ums4tm1}
				\end{eqnarray}
where $c_{\theta}=\cos\theta$ and $s_{\theta}=\sin\theta$. The above matrix has the TM$_1$ mixing structure (including the phase factor $\gamma$) mentioned in Eq. (\ref{utm}). This is also an example of the method of semi-direct approach to the flavor model building. Similarly, the generic structure for the structure for TM$_2$ mixing matrix can be written as \begin{eqnarray}
				U_{{\rm TM}_2}=	\left(
					\begin{array}{ccc}
						\frac{2c_{\theta}}{\sqrt{6}} &\frac{1}{\sqrt{3}} &\frac{2s_{\theta}}{\sqrt{6}}e^{-i\gamma}\\
			-\frac{c_{\theta}}{\sqrt{6}}+\frac{s}{\sqrt{2}}e^{i\gamma} &\frac{1}{\sqrt{3}}  &-\frac{s_{\theta}}{\sqrt{3}}e^{-i\gamma}-\frac{c_{\theta}}{\sqrt{2}}\\
				-\frac{c_{\theta}}{\sqrt{6}}+\frac{s}{\sqrt{2}}e^{i\gamma} &\frac{1}{\sqrt{3}} &-\frac{s_{\theta}}{\sqrt{3}}e^{-i\gamma}+\frac{c_{\theta}}{\sqrt{2}}
					\end{array}
					\right)\label{ums4tm2}. 
				\end{eqnarray}
Explicit models to obtain TM$_1$ and TM$_2$ mixing can be found in~\cite{Shimizu:2011xg, King:2011zj, Luhn:2013lkn, deMedeirosVarzielas:2012apl}. The above discussion shows special cases of the TBM mixing which can still be relevant for models with discrete flavor symmetries. Now imposing sufficient corrections to the other fixed mixing schemes like BM, GR, HG we can make them consistent with observed data. In this case, the modified mixing matrix can be obtained by lowering the residual symmetry $G_{\nu}$ for the neutrino sector. This generates a correction matrix to these fixed mixing patterns. The general form of these corrections can be summarized as~\cite{Petcov:2017ggy,Hernandez:2013vya} \begin{equation}
    U_{PMNS}= U_{e}^{\dagger} U_{\psi} U_{0} U_{p},
\end{equation}
where $U_0$ is the general form of the relevant fixed pattern mixing scheme mentioned in Eq. (\ref{ugen}), $U_{e}$ is the generic correction matrix  and $U_{\psi}, U_{p}$ are additional phase matrices.  These corrections help us to obtain interesting correlations among $\sin\theta_{12}, \sin\theta_{23}, \sin\theta_{13}$ and $\delta_{CP}$  of PMNS mixing matrix~\cite{Petcov:2014laa}. Below in Tab.~\ref{tab:tm12prediction} we mention the typical predictions for TM$_1$ and TM$_2$ mixing matrices. 
\begin{table}[h]
    \centering
    \begin{tabular}{|c|c|c|}
    \hline
          & TM$_1$ & TM$_2$ \\
    \hline      
        $|U_{e2}|$ & $\left|\frac{\cos\theta}{\sqrt{3}}\right|$ & $\frac{1}{\sqrt{3}}$\\
    \hline     
    $|U_{e3}|$ & $\left|\frac{\sin\theta}{\sqrt{3}} e^{-i\gamma}\right|$ & $\left|\frac{2\sin\theta}{\sqrt{6}} e^{-i\gamma}\right|$\\
    \hline    
    $|U_{\mu 3}|$ & $\left|\frac{\cos\theta}{\sqrt{2}} + \frac{\sin\theta}{\sqrt{3}} e^{-i\gamma}\right|$ &  $\left|-\frac{\cos\theta}{\sqrt{2}} - \frac{\sin\theta}{\sqrt{6}} e^{-i\gamma}\right|$\\
    \hline     
    $\sin^2\theta_{12}$ & $1-\frac{2}{3-\sin^2\theta}$ & $\frac{1}{3-2\sin^2\theta}$\\
    \hline     
    $\sin^2\theta_{13}$ & $\frac{1}{3}\sin^2\theta$ & $\frac{2}{3}\sin^2\theta$\\
    \hline   
    $\sin^2\theta_{12}$ & $\frac{1}{2} \left( 1- \frac{\sqrt{6}\sin 2 \theta \cos\gamma}{3-\sin^2\theta} \right)$ & $\frac{1}{2} \left( 1+ \frac{\sqrt{3}\sin 2 \theta \cos\gamma}{3-\sin^2\theta} \right)$\\
    \hline    
    $J_{CP}$ & $-\frac{1}{6\sqrt{6}}\sin2\theta \sin\gamma$
 &  $-\frac{1}{6\sqrt{3}}\sin2\theta \sin\gamma$
 \\
    \hline 
$\sin\delta_{CP}$ & -$\frac{(5+\cos2\theta)\sin\gamma}{\sqrt{(5+\cos2\theta)^2-24 \sin^22\theta\cos^2\gamma}}$ & -$\frac{(2+\cos2\theta)\sin\gamma}{\sqrt{(2+\cos2\theta)^2-3 \sin^22\theta\cos^2\gamma}}$\\
\hline
    \end{tabular}
    \caption{Mixing parameters in the TM$_1$ and TM$_2$ scenarios including the Jarlskog invariant $J_{CP}$.}
    \label{tab:tm12prediction}
\end{table}

As mentioned earlier, the mixing matrix in Eq. (\ref{ugen}) and the corresponding mass matrix   Eq. (\ref{mmutau}) obeys the underlying $\mu-\tau$ symmetry (also known as the $\mu-\tau$ permutation symmetry). As this is outdated now for obvious reason, there is another class of flavor CP model known as $\mu-\tau$ reflection symmetry~\cite{Xing:2015fdg}. This can be expressed  as the transformation : 
\begin{equation}\label{eq:reflection}
    \nu_{e} \rightarrow \nu_{e}^C, ~\nu_{\mu} \rightarrow \nu_{\tau}^C,~ \nu_{\tau} \rightarrow 
\nu_{\mu}^C
\end{equation}
where `C' stands for the charge conjugation of the corresponding neutrino field, under which the neutrino mass term remains unchanged.  These leads to the predictions $\theta_{23}= 45^\circ$, $\delta_{CP}= 90^\circ ~{\rm or}~ 270^{\circ}$. Clearly, such mixing scheme is still experimentally viable~\cite{Esteban:2020cvm}.  Under such symmetry the elements of
lepton mixing matrix satisfy :
\begin{equation}\label{umutau}
 |U_{\mu i}|=|U_{\tau i}|~~~~~{\rm where}~~~{i=1,2,3}. 
\end{equation}
Such mixing scheme is also known as cobimaximal (CBM) mixing scheme~\cite{Grimus:2017itg}. Eq. (\ref{umutau}) indicates that the moduli of $\mu$ and $\tau$  flavor elements of the $3\times 3$ neutrino mixing matrix are equal. 
With these constraints, the neutrino mixing matrix can be parametrized as~\cite{Harrison:2002et,Grimus:2003yn}
\begin{eqnarray}\label{u3b3}
U_{0}&=&\left(
\begin{array}{ccc}
 u_1  & u_2   & u_3 \\
v_1  & v_2   & v_3 \\
v^*_1  & v^*_2   & v^*_3 
\end{array}
\right), 
\end{eqnarray}
where the entries in the first row, $u_i$'s are real (and non-negative) with trivial values of the Majorana  phases. Here $v_i$  satisfy the orthogonality condition 
${\rm Re}(v_jv_k^*)=\delta_{jk}-u_k u_k$.  In~\cite{Grimus:2003yn}, it was argued that the mass matrix leading to  the mixing matrix given in Eq.(\ref{u3b3}) 
can be written as 
\begin{eqnarray}
{M}_0&=&\left( \label{m3b3}
\begin{array}{ccc}
 a  & d   & d^* \\
d  & c   & b \\
d^* & b   & c^*
\end{array}
\right), 
\end{eqnarray}
where $a,b$ are real and $d,c$ are complex parameters. As a consequence of the symmetry given in Eq.(\ref{umutau})-(\ref{m3b3}),
we obtain the  predictions for maximal $\theta_{23}=45^\circ$ and $\delta= 90^\circ ~{\rm or}~ 270^{\circ}$ in the basis where the 
charged leptons are considered to be diagonal. This scheme however still leaves room for nonzero $\theta_{13}$. Realization of such a mixing pattern is possible with various discrete flavor symmetries ($A_4, \Delta(27)$, etc.), for example see \cite{Chakraborty:2019rjc,CarcamoHernandez:2020udg,Vien:2020dlk,Vien:2021bsv, Karmakar:2022cbm}.

\subsection{Flavor and CP symmetries}
The transformations given in Eq. (\ref{eq:reflection}), represents an interesting extension of the discrete symmetry framework discussed earlier with an additional invariance under CP transformations. Along with residual symmetries, such transformations also lead to new invariance conditions on the mass matrices. For example, the mass matrix given in Eq. (\ref{m3b3}) is invariant under  
\begin{equation}
    \mathcal{S}^T {M}_0 \mathcal{S} = {M}^*_0, 
\end{equation}
where the transformation matrix is given by 
\begin{eqnarray}
 \mathcal{S}&=&\left( \label{S}
\begin{array}{ccc}
 1  & 0   & 0 \\
0  & 0   & 1 \\
0 & 1   & 0
\end{array}
\right)
\end{eqnarray}
and such transformations are usually referred to as generalized CP symmetry transformation. The existence of both discrete flavor and generalized CP symmetries determines the possible structure of the generalized CP symmetry matrices. For example, if we consider a discrete flavor symmetry $G_f$ in the lepton sector, the transformation matrix given in Eq. (\ref{S}) satisfy the {\it consistency condition} given by~\cite{Feruglio:2012cw, Holthausen:2012dk}
\begin{equation}
     \mathcal{S} \rho(g)^*  \mathcal{S}^{-1} =\rho (u(g)). 
\end{equation}
where $u$ is an automorphism of a group of $G_f$ which maps an element $g \in G_f$ into $g'=u(g)\in G_f$ where the latter belong to the conjugacy class of $g^{-1}$. 
One can derive this condition by applying a generalized CP symmetry transformation, subsequently, a flavor transformation associated with the group element $g=G_{f}$ and an inverse generalized CP symmetry transformation in order. As the Lagrangian remains unchanged the resulting transformation must correspond to an element of $G_f$.  This can lead to interesting predictions for the leptonic CP phases (both Dirac and Majorana) and the mixing angles. Few such examples are listed here for generalized CP symmetry transformation with various discrete groups such as  $\Delta(27)$~\cite{ Nishi:2013jqa}, $S_4$~\cite{Feruglio:2013hia,Penedo:2018gpg,  Li:2013jya, Li:2014eia,Penedo:2017vtf} $A_4$~\cite{Ding:2013bpa}, $\Delta(48)$~\cite{Ding:2013nsa,Ding:2014hva}, $\Delta(6n^2)$~\cite{King:2014rwa,Hagedorn:2014wha}, $\Delta(96)$~\cite{Ding:2014ssa}, $A_5$~\cite{Feruglio:2015jfa, DiIura:2018fnk}, $\Delta(3n^3)$~\cite{Ding:2015rwa}.

\subsection{Higher Order Discrete Groups}
In the above we have discussed various fixed mixed schemes (BM, TBM, GR, HG) which are ruled out   by data and few possible mixing schemes (TM$_{1}$,TM$_{2}$,CBM) mixing schemes which are still consistent with observations.   In this regard we have also seen that smaller discrete groups (such as $A_4, S_4, \Delta(27), T_7$, etc.) still can explain the correct mixing with 'appropriate  adjustments' in the old flvour symmetric models. Now we will discuss few aspects of explaining lepton mixing with larger groups. In this technique we   look for new groups that predict a different type of leptonic mixing pattern. Therefore,  our new method is to start with much higher order groups ($G_f$) which which essentially breaks down to two groups $G_e$ and $G_{\nu}$ for charged leptons and neutrino sector.   For example, when we start 
    \begin{figure}[h!]
\includegraphics[scale=.55]{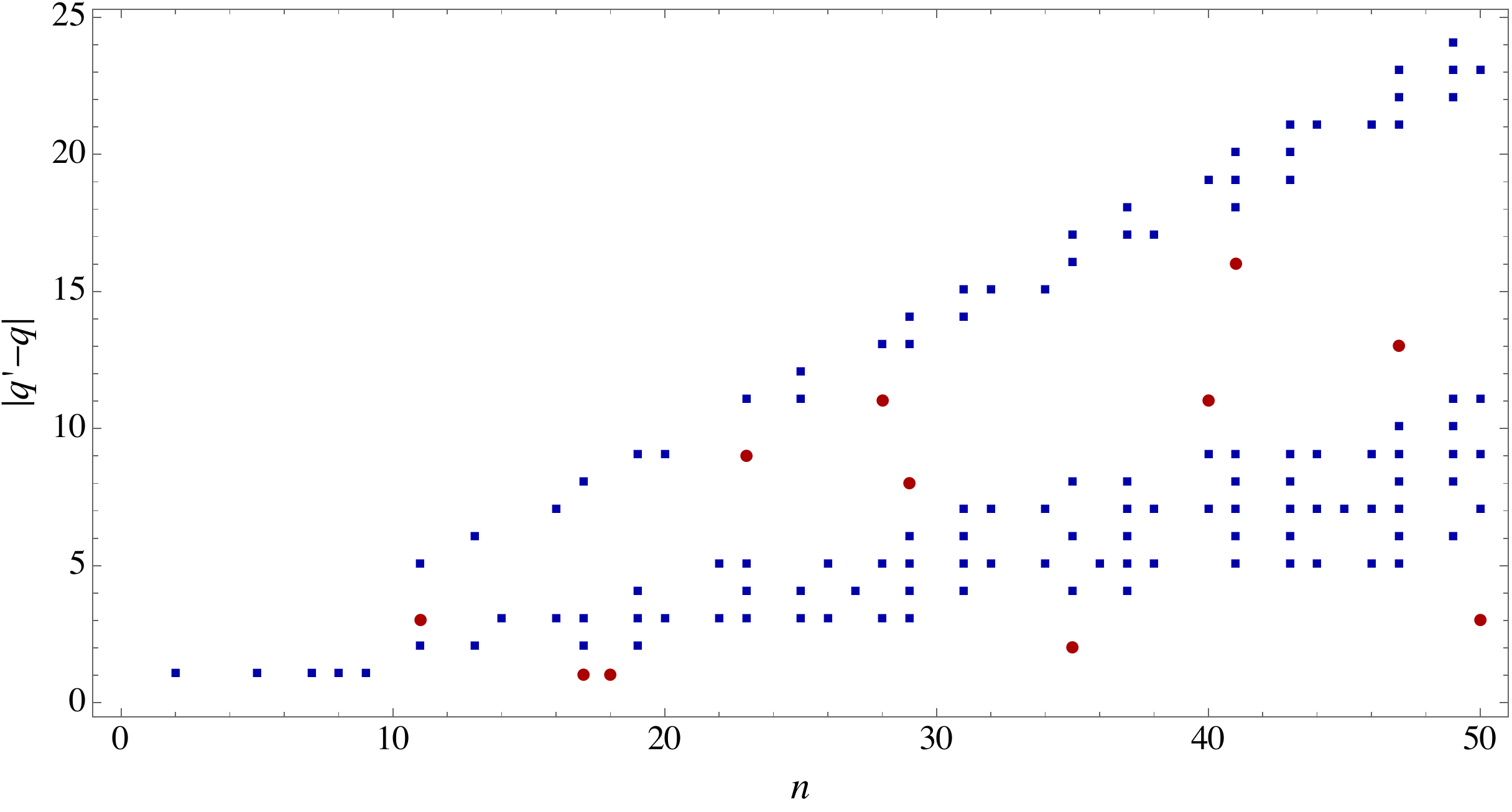}
\caption{The values of $n \le 50$ and $|q'-q|$ ($q, q' = 0, 1, ..., n-1$) leading to the viable columns of leptonic mixing matrix. The blue squares (red dots) indicate that the corresponding prediction is consistent with the first (third) column of $U_{\rm PMNS}$ matrix within $3\sigma$. Each point represents a unique solution obtained by the smallest possible values of $n$ and $|q'-q|$. Figure taken from \cite{Joshipura:2016quv}, Creative Commons CC BY license.}
\label{fig:ketan}       
\end{figure}
with a discrete group of the order less than 1536, and the residual symmetries are fixed at $G_e=Z_3$ and $G_{\nu}=Z_2 \times Z_3$, the only surviving group which can correct neutrino mixing are $\Delta(6 n^2)$ with $n=10$,   $(Z_{18} \times Z_6)\times S_3$ and $\Delta(6 n^2)$ with $n=16$~\cite{Holthausen:2012wt}. In a more updated study, considering discrete subgroups of $U(3)$, it has been shown that the smallest group which satisfy 3$\sigma$ allowed range of neutrino oscillation data is $\Delta(6n^2)$ with $n=18$, i.e., the order of the group is 1944~\cite{Joshipura:2016quv}. This study~\cite{Joshipura:2016quv} also proposes the a analytical formula to predict the   columns of lepton mixing matrix  as given in Fig. \ref{fig:ketan}. For a review of similar studies, the reader are referred to~\cite{Yao:2015dwa} and references therein.

\subsection{Flavor symmetry and Grand Unified Theory}
To have a complete understanding of the flavor problem in particle physics, propositions are there to combine  family  symmetry and Grand
Unified Theory (GUT)~\cite{Georgi:1974sy, Pati:1974yy, Fritzsch:1974nn}. This 
Grand Unified Theories of Flavor therefore has potential to explain observed lepton mixing due to the presence of the discrete flavor symmetry as well as the theory also include GUT predictions through  the associated Clebsch factors~\cite{Georgi:1979df, Antusch:2013rxa}. This can provide a novel connection between the smallest lepton mixing angle, i.e., the reactor mixing angle  $\theta_{13}$ and the largest quark mixing angle, the Cabibbo angle $\theta_C$, which are related by~\cite{Antusch:2013kna, Antusch:2012fb, Marzocca:2011dh}
\begin{equation}
    \theta_{13}=\theta_C / \sqrt{2}. 
\end{equation}
Several  combinations of discrete flavor symmetry and GUT are possible based on their choice and for classification of such models the readers are referred to the review~\cite{King:2017guk} and references therein for explicit models. These models however greatly depends on the  symmetry breaking pattern and  vacuum alignment of the fields. In addition, such unified constructions have consequences in leptogenesis~\cite{Bjorkeroth:2016lzs, deAnda:2017yeb,Gehrlein:2015dxa,Gehrlein:2015dza, deAnda:2018oik} and at LHC~\cite{Belyaev:2018vkl}. Recently, GUT frameworks have also been implemented in modular invariance approach~\cite{Charalampous:2021gmf, Ding:2021eva, Ding:2021zbg, King:2021fhl,Chen:2021zty}. 


\subsection{Flavor Symmetry and Multi-Higgs Doublet Models}

Imposition of a flavor symmetry on the leptonic part of the Yukawa Lagrangian has been discussed in  \cite{Lam:2007ev,King:2017guk, Altarelli:2010gt}. In the SM, application of family symmetry is limited due to the Schur's lemma \cite{Chaber:2018cbi} which implies that for three-dimensional mass matrices of charged leptons and neutrinos their diagonalization matrices are proportional to identity, thus the \texttt{PMNS} matrix becomes trivial. This drawback can be overcome in two ways. One approach is to break the family symmetry group by a scalar singlet, so-called flavon \cite{King:2013eh}. Despite many attempts it has failed to reconstruct the \texttt{PMNS} matrix. In the second approach a non-trivial mixing can be  achieved by extending the Higgs sector by additional multiplets \cite{Machado:2010uc,deMedeirosVarzielas:2021zqs, Varzielas:2022lye}. In this context, basing on numerical approach, multi Higgs doublet models were examined which include two Higgs doublets  (2HDM) \cite{Chaber:2018cbi} and three Higgs doublets (3HDM) \cite{vergeest:2022mqm}.  Finite, non-Abelian subgroups of the $U(3)$ group up to the order of 1035 were investigated. The scan over all finite discrete groups  was performed with the help of the computer algebra system gap \cite{PhysRevD.92.096010}. {To significantly reduce enormous number of evaluated subgroups a limitation to 
subgroups with irreducible three dimensional faithful representations has been performed. For both 2HDM and 3HDM models, analyses were made assuming that neutrinos are Dirac or Majorana particles. From the model building point it was also assumed that the total Lagrangian has a full flavor symmetry and that the Higgs potential is only invariant in form \cite{Branco:2011iw,Davidson:2005cw}. }{\it Results were utterly negative for 2HDM { - for such an extension of the SM there is no discrete family symmetry that would fully clarify the masses and parameters of the mixing matrix for leptons.}} However, 3HDM provides nontrivial relations among the lepton masses and mixing angles which leads to nontrivial results. Namely, {\it some of the scanned groups provide either the correct neutrino masses and mixing angles
or the correct masses of the charged leptons}. The group $\Delta(96)$ is the smallest group compatible with the
experimental data of neutrino masses and PMNS mixing whereas $S_4$ is an approximate symmetry of Dirac
neutrino mixing, with parameters staying about 3{$\sigma$} apart from the measured $\theta_{12}, \theta_{23}, \theta_{13}$, and $\delta_{CP}$.
Thus, investigations beyond 3HDM are worth future studies where flavor symmetries of the entire lepton sector can be realized. In parallel, using the same methodology, we propose to  investigate the horizontal symmetry with the full $6 \time 6$ neutrino mixing matrix. In this way a natural link with 3+N Majorana neutrinos and CP-violating scenarios can be studied.

\subsection{Flavor Symmetry and neutrino mass models}

Apart from the observed pattern of neutrino mixing, the origin of tiny neutrino mass is still unknown to us. One of the most popular ways to generate tiny neutrino mass is to start with the high scale suppressed lepton number violating Weinberg operator $H L^T c LH $ mentioned earlier, which is non-renormalizable: $c$ is the dimensional operator $[mass]^{-1}$. This can give rise to various 
seesaw mechanisms~\cite{Minkowski:1977sc, Mohapatra:1979ia, Yanagida:1979as, GellMann:1980vs, Glashow:1979nm, Schechter:1980gr, deGouvea:2016qpx} such as type-I, type-II, type-III, inverse and linear seesaw.  Examples for discrete flavor symmetric models for type-I, type-II, and inverse seesaw which are very efficient in explaining tiny mass as well as correct  mixing as observed by the neutrino oscillation experiments can be found in~\cite{Karmakar:2014dva,Karmakar:2015jza,Karmakar:2016cvb,Datta:2021zzf}. 
{Another class of models} that can be connected with flavor symmetries are radiative neutrino mass models in which masses of the neutrinos are absent at the tree-level and are generated at 1- or higher-loop orders. These models provide an explanation for the lightness of neutrino masses. Most of the radiative models involve Majorana neutrinos although the Dirac type of neutrinos is also possible. A broad review of various radiative neutrino models can be found in \cite{Cai:2017jrq}. The key feature of these models is that they can be verified experimentally because the masses of exotic particles which take part in the mass generation are in $\mathcal{O}$(1-100) TeV range. On top of that, there are radiative neutrino mass models on the market with various additional flavor symmetries, for instance, modular $S_3$ and $A_4$ \cite{Okada:2021aoi,Nomura:2022hxs, Nomura:2021pld,Otsuka:2022rak,Okada:2019xqk}. Furthermore, these models provide extra contributions to neutrinoless double beta decay, electric dipole moments, anomalous magnetic moments and meson decays as well as matter-antimatter asymmetry, and may also solve the dark matter problem. 

While neutrino oscillation experiments are insensitive to the nature of neutrinos, experiments looking for lepton number violating
signatures can probe the Majorana nature of neutrinos. Neutrinoless double beta decay is one such lepton number violating process which has been searched for at several experiments without any positive result so far but giving stricter bounds on the effective neutrino mass. Although negative results at neutrinoless double beta decay experiments do not prove that the light neutrinos are of Dirac nature, it is nevertheless suggestive enough to come up with scenarios predicting Dirac neutrinos with correct mass and mixing. There have been several
proposals already that can generate tiny Dirac neutrino masses~\cite{Babu:1988yq,Peltoniemi:1992ss,Aranda:2013gga, Ma:2015mjd, CentellesChulia:2016rms, Abbas:2016qbl}.  In~\cite{Borah:2017dmk,Borah:2018nvu,CentellesChulia:2018gwr,CentellesChulia:2018bkz}, the authors showed that it is possible to propose various seesaw mechanism (type-I, inverse and linear seesaw) for Dirac neutrinos with $A_4$ discrete flavor symmetry. Here the symmetry is chosen in such a way that it naturally explains the hierarchy among different terms in the neutrino mass matrix, contrary to the conventional seesaws where this hierarchy is ad-hoc.

\section{Implications of Flavor Symmetry in Various Frontiers}
 
Flavor symmetry can be the potentially important link to connect the outstanding puzzle of neutrino masses and mixings to the various aspect of particle, cosmology, and astroparticle physics such as neutrinoless double-beta decay, lepton flavor violating decays,  the nature of dark matter,  observed baryon asymmetry of the Universe, nonstandard interactions, etc. If these seemingly uncorrelated sectors are connected by a flavor symmetry,  the constraints from various cosmological, collider, and neutrino experiments may probe the existence of such symmetry.   This is one of the biggest challenges to the studies with discrete flavor symmetry.    

\subsection{Flavor symmetry and dark matter}

Various astrophysical evidence such as the large-scale structure of the Universe, gravitational lensing, and rotation curve of galaxies supports the postulate of the existence of dark matter~\cite{Bertone:2004pz}, however, a laboratory discovery is still awaited. The relic abundance of dark matter has been measured by WMAP~\cite{WMAP:2012nax} and PLANCK~\cite{Planck:2018nkj} satellite experiments to be about $26.8\%$ of the total energy budget of the Universe. This hints toward a broad classification of dark matter scenarios such as weakly interacting massive particle
(WIMP), feebly interacting massive particle (FIMP), strongly interacting massive particle (SIMP), asymmetric dark matter (ADM), for review of various dark matter candidates see~\cite{Feng:2010gw, Bernal:2017kxu, Hall:2009bx, Bernal:2018ins, Petraki:2013wwa}. Over the years, several attempts have been made to connect neutrino physics with dark matter~\cite{Mohapatra:2002ug, Ma:2006km, Boehm:2006mi, Hambye:2006zn, Ma:2006fn, Allahverdi:2007wt, Sahu:2008aw, Gu:2007ug, Bhattacharya:2018ljs, Chianese:2020khl,  Chianese:2018dsz,Bhattacharya:2021jli}. In a toy example~\cite{Bhattacharya:2018ljs}, the authors showed a one-to-one correspondence with WIMP dark matter and type-I seesaw. Earlier we have discussed that discrete flavor symmetric constructions have the potential to explain neutrino masses and mixing as well as can ensure the stability of dark matter as we proceed.  

\begin{figure}
    \centering
    \includegraphics[width=0.5\textwidth]{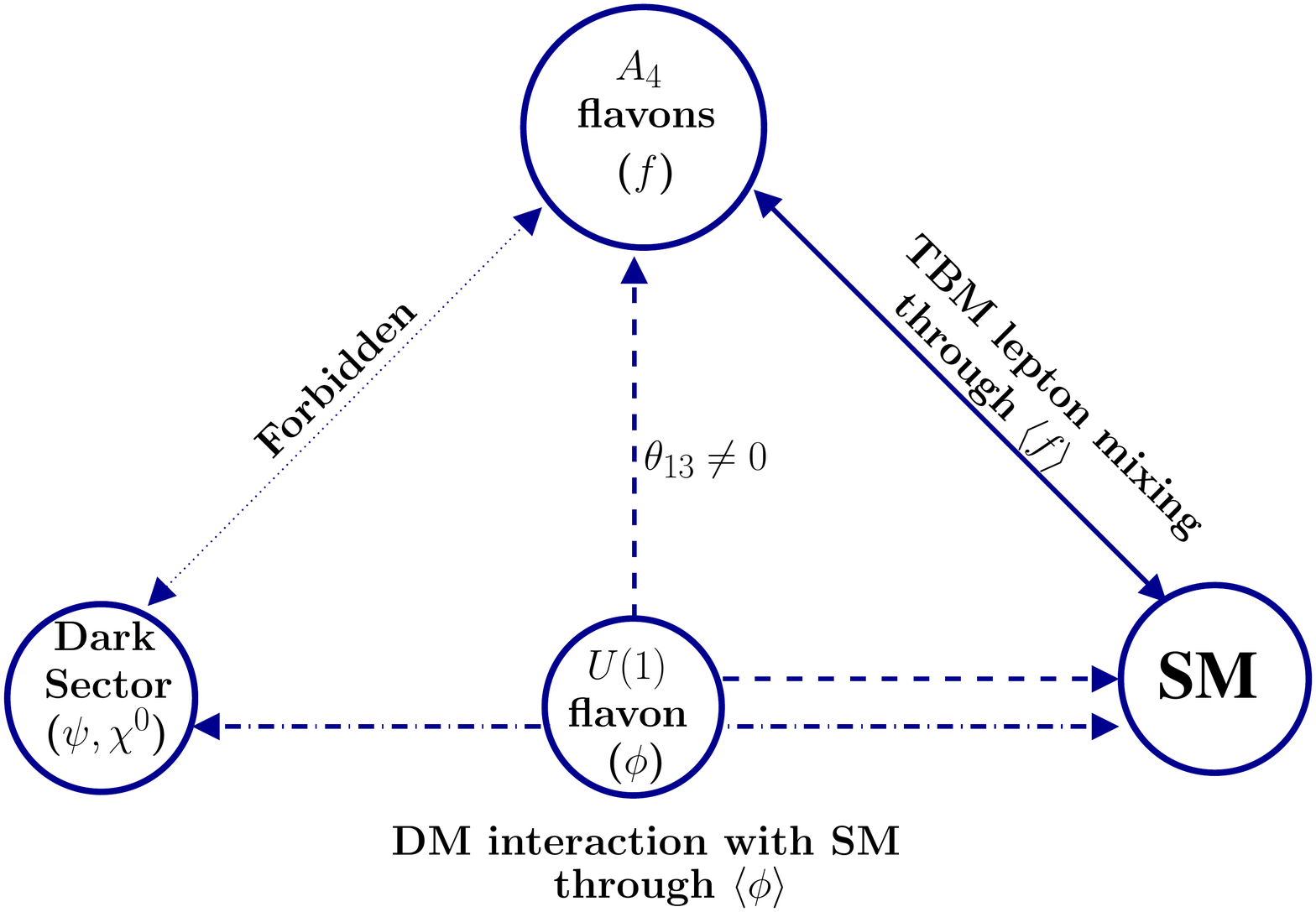}
    \caption{A schematic representation of dark matter ($\psi, \chi^0$) interaction with SM to generate non-zero $\theta_{13}$ in the presence of the $U(1)$ flavor symmetry. The $A_4$ flavons help in generating base TBM mixing. Figure adapted from~\cite{Bhattacharya:2016lts} under CC BY 4.0 license.}
    \label{fig:dmscm}
\end{figure}

For example~\cite{Bhattacharya:2016lts,Bhattacharya:2016rqj}, with vectorlike singlet ($\chi^0$)-doublet ($\psi$) dark matter particle spectrum assisted with additional scalars such as ($\phi,\eta$ charged under a global $U(1)$ flavor symmetry) the interaction between dark matter and neutrino sector can be written as 
\begin{equation}
\mathcal{L}_{int}=\left(\frac{\phi}{\Lambda}\right)^n \bar{\psi} \Tilde{H}\chi^0+\frac{(HL^TLH)\phi\eta}{\Lambda^3}.   \label{eq:dmint1}
\end{equation}
The first term in Eq. (\ref{eq:dmint1})  having a Yukawa-like configuration, acts like a Higgs portal coupling of the dark matter potentially accessible at various ongoing and future direct search and collider  experiments.  The second term in Eq. (\ref{eq:dmint1}) plays a crucial role in generating non-zero $\theta_{13}$ as the existing $A_4$ flavons of theory ensures TBM mixing. A schematic view for this is given in Fig.~\ref{fig:dmscm}. 
Clearly, once the $U(1)$ flavor symmetry is broken, it ensures both stability of dark matter as well as generates non-zero $\theta_{13}$. The coupling strength of the dark matter $\left(\frac{\phi}{\Lambda}\right)^n = \epsilon^n$ is constrained by the correct dark matter relic abundance and $\epsilon$ is proportional to the magnitude of $\theta_{13}$. Additionally, the future precise measurement of the leptonic CP phase by T2K and NO$\nu$A experiments will reduce the uncertainty in $n$~\cite{Bhattacharya:2016rqj}. Few other examples of studies of  dark matter   with discrete flavor symmetry can be found in~\cite{Esteves:2010sh, deMedeirosVarzielas:2015ybd, CarcamoHernandez:2013yiy, Hirsch:2010ru, Boucenna:2011tj, Verma:2021koo, Kajiyama:2010sb, Mukherjee:2015axj, Arcadi:2020zai, CentellesChulia:2016fxr,Mukherjee:2017pzq,Gautam:2019pce,deMedeirosVarzielas:2015lmh}. 

Apart from WIMP dark matter paradigm, discrete flavor symmetric constructions can also be extended in the freeze-in mechanism of dark matter. In such a scenario, non-thermal dark matter populating the Universe via freeze-in
mechanism requires tiny dimensionless couplings ($\sim 10^{12}$). On the other hand, if neutrinos have a tiny Dirac mass, it also requires a coupling of a similar order of magnitude. In~\cite{Borah:2018gjk} the authors have shown that such tiny coupling required in both dark and visible sectors may have its origin in $A_4$ discrete flavor symmetry. In another effort~\cite{Borah:2019ldn} it has been shown that the non-zero value of the reactor mixing angle $\theta_{13}$ can originate from a Planck scale suppress operators and as well as realize super-WIMP dark matter scenario. The study of dark matter in the context of non-Abelian discrete symmetries demands a lot more attention to explore the connection of neutrino physics and dark matter, if any.  

\subsection{Flavor symmetry and baryon asymmetry of the Universe (BAU)}

An attractive mechanism for BAU production is leptogenesis~\cite{Fukugita:1986hr}. 
The presence of various seesaw realizations~\cite{Minkowski:1977sc, Mohapatra:1979ia, Yanagida:1979as, GellMann:1980vs, Glashow:1979nm, Schechter:1980gr} of light neutrino
masses in the models with discrete flavor symmetry enables us to study leptogenesis through the decay of such heavy degrees of freedom. For example, in the case of the simple type-I seesaw, lepton asymmetry can be elegantly generated through the out-of-equilibrium decay
of heavy right-handed neutrinos (RHNs) in the early Universe. Through electroweak sphalerons, this lepton asymmetry is converted to a net baryon asymmetry~\cite{Kuzmin:1985mm}. The CP asymmetry parameter $\epsilon^\alpha_i$ can be evaluated from  the interference between the tree and one-loop level decay amplitudes of RHN, $N_i$ decaying into a lepton doublet $L_\alpha$ with specific flavor $\alpha$ and scalar Higgs ($H$), and can be written as follows~\cite{Fukugita:1986hr, Covi:1996wh},  
\begin{align}
	\epsilon_{i}^{\alpha}=& \frac{\Gamma(N_i \to L_{\alpha} H )-\Gamma(N_i \to \overline{L}_{\alpha} \bar{H} )}{\Gamma(N_i \to L_{\alpha} H )+\Gamma(N_i \to \overline{L}_{\alpha} \bar{H} )}
	=\frac{1}{8\pi}\sum_{j\ne{i}}\frac{{\rm Im}\left[\left((\hat{Y}^{\dagger}_{D}\hat{Y}_{D})_{ij}\right)^{2}\right]}
	{(\hat{Y}^{\dagger}_{D}\hat{Y}_{D})_{ii}}f\left(\frac{M^2_{j}}{M^2_{i}}\right), \label{eq:cpasym} 
\end{align}
where the loop factor $f(x)$ in the above expression is
\begin{equation}
	f(x)=\sqrt{x}\left[\frac{2-x}{1-x}-(1-x)\ln\left(1+\frac{1}{x}\right)\right] 
\end{equation}
with $x=M^2_{j}/M^2_{i}$. For reviews and various aspects of leptogenesis see~\cite{Davidson:2008bu, Pilaftsis:2009pk, Dev:2017trv, Chun:2017spz}. Here $\hat{Y}_{D}$ is the Dirac Yukawa coupling in the basis where both RHNs and charged leptons are diagonal. As mentioned earlier, in models with discrete flavor symmetry,  structures of the Dirac Yukawa coupling, charged lepton, and RHN mass matrices are entirely instrumented by the associated symmetry.   Hence the involved discrete symmetry may have a significant impact on neutrino masses, mixing, and Dirac CP phase as well high energy CP violation required for leptogenesis. As TBM mixing was a potential candidate for lepton mixing matrix several analyses have been performed to explain TBM mixing and leptogenesis~\cite{Jenkins:2008rb,Hagedorn:2009jy, Branco:2009by}. Interestingly, it was also observed~\cite{Jenkins:2008rb} that in the exact
TBM limit, the CP asymmetry vanishes since  $\hat{Y}^{\dagger}_{D}\hat{Y}_{D} \propto \mathbb{I}$. Therefore in the exact TBM scenario leptogenesis was realized either by introducing higher correction in the Dirac Yukawa~\cite{Jenkins:2008rb, Hagedorn:2009jy} or by introducing renormalization group effects~\cite{Branco:2009by}. Now,  in the precision era of the neutrino mixing parameters it is essential to study the effect of non-zero $\theta_{13}$ and CP phases. In the $A_4$ type-I seesaw scenario~\cite{Karmakar:2014dva}, the authors showed that a minimal (by adding a non-trivial $A_4$ singlet) modification to the existing Altarelli-Feruglio~\cite{Altarelli:2005yx} model may give rise to non-zero $\theta_{13}$ as well as the effect in generating observed BAU. Here the high energy CP phases or the Majorana phases appearing in the CP asymmetry also get constrained by the low energy neutrino oscillation data which are otherwise insensitive to the oscillation experiments. In a more minimal study~\cite{Datta:2021zzf} the authors showed that with the presence of only three $A_4$ flavons both non-zero $\theta_{13}$ and BAU (incorporating renormalization group effects) can be generated. In order to unify all the sources of CP violation in the theory,  the CP symmetry can be spontaneously
broken by the VEV of a singlet field where its magnitude generates non-zero $\theta_{13}$ and phase factor becomes directly proportional to the CP asymmetry parameter in general $A_4$ type-I+II scenario~\cite{Karmakar:2015jza}. Recently, it has also been shown that BAU can be successfully generated in a  $S_4$ discrete symmetric framework with TM$_1$ mixing \cite{Chakraborty:2020gqc} taking lepton flavor effects into consideration. For a few more recent studies  on various discrete flavor symmetry and leptogenesis see~\cite{Boruah:2021ayj, Sarma:2021icl, Gautam:2020wsd, Das:2019ntw, Borah:2017qdu,Bjorkeroth:2016lzs,Bjorkeroth:2015tsa}.

\subsection{Flavor symmetry and collider physics with $\Delta (6n^2)$ }

The same complex Yukawa interactions ($Y_D$) of the RHNs with the SM lepton and Higgs doublets involved in leptogenesis, in addition with the Majorana masses of the RHNs are also responsible for the light neutrino masses via the type-I seesaw mechanism~\cite{Minkowski:1977sc, Mohapatra:1979ia, Yanagida:1979as, GellMann:1980vs, Glashow:1979nm, Schechter:1980gr}. The high-energy CP phases present in $Y_D$ that are responsible for leptogenesis are in general unrelated to the low-energy CP phases in $U_{PMNS}$. Since the experiments are only sensitive to the low-energy CP phases, by incorporating residual flavor and CP symmetries the high- and low-energy CP phases can be related. This can be achieved by introducing a non-abelian discrete flavor symmetry $G_f$ combined with a CP symmetry, both acting non-trivially on the flavor space. Since in this case the PMNS mixing matrix depends on a single free parameter, this turns out to be highly constraining and predictive for both low- and high-energy CP phases as well as the lepton mixing angles ~\cite{Feruglio:2012cw, Holthausen:2012dk, Chen:2014tpa}. In this section, we will discuss this setup with $G_f$ being a member of the series of groups $\Delta(6\, n^2)$~\cite{Escobar:2008vc}, known to give several interesting neutrino mixing patterns~\cite{King:2014rwa, Hagedorn:2014wha, Ding:2014ora, Ding:2015rwa}.  

This framework provides an excellent probe of flavor symmetries in experiments at colliders. It also leads to points of {\it enhanced residual symmetry} (ERS) at which one of the three RHNs becomes long-lived and can be probed through long-lived particle (LLP) searches~\cite{Alimena:2019zri, Alimena:2021mdu}, while the remaining two RHNs can be searched via prompt/displaced vertex signals at the LHC~\cite{Deppisch:2015qwa, Cai:2017mow} and future hadron collider FCC-hh \cite{FCC:2018byv,FCC:2018vvp}, see Fig.~\ref{fig:collider}.


The discrete groups $\Delta (6 \, n^2)$, $n \geq 2$ integer~\cite{Escobar:2008vc} can be
described in terms of three generators $a$, $b$, $c$ and $d$ fulfilling the relations
\begin{equation*}
a^3 \, = \, e \; , \;\; c^n \, = \, e \; , \;\; d^n \, = \, e \; , \;\; c \, d \, = \, d \, c\; , \;\; a \, c \, a^{-1} \, = \, c^{-1} d^{-1} \; , \;\; a \, d \, a^{-1} \, = \, c \, ,
\end{equation*}
\begin{equation}
b^2 \, = \, e \; , \;\; (a \, b)^2\, = \, e \; , \;\; b \, c \, b^{-1} \, = \, d^{-1} \; , \;\; b \, d \, b^{-1} \, = \, c^{-1} \, .
\end{equation}
with $e$ being the identity element of the group.

After breaking down of the flavor group $G_f$ at low-energies, we have residual flavor and CP symmetries in the neutrino and charged lepton sector. The residual symmetry in the charged lepton sector is chosen to be the diagonal abelian subgroup of $Z_3$ i.e. $G_\ell=Z_3^{(\mathrm{D})}$. While in the neutrino sector, we chose the residual symmetry to be $G_\nu=Z_2 \times {\rm CP}$. Since the generators of  $Z_2$ symmetry  $Z (\mathrm{\mathbf{r}})$  and CP symmetry  $X (\mathrm{\mathbf{r}})$ commute they fulfill
\begin{equation}
X (\mathrm{\mathbf{r}}) \, Z (\mathrm{\mathbf{r}}) - Z (\mathrm{\mathbf{r}})^\star \, X (\mathrm{\mathbf{r}}) \, = \, 0
\end{equation}
The CP symmetry transformations corresponds to the automorphisms of the flavor group. The mismatch of the residual symmetries $G_\ell$ and $G_\nu$ determines the form of lepton mixing.
The residual symmetries $G_\ell$ and $G_\nu$ determine the forms of the lepton mixing matrix, charged lepton mass matrix, the neutrino Yukawa $Y_D$ and the RHN Majorana mass matrix $M_R$. Since in our chosen basis, the charged lepton mass matrix is diagonal, the charged lepton sector does not contribute to the lepton mixing. As for the neutrino sector, we assume the neutrino Yukawa coupling matrix $Y_D$ to be invariant under $G_\nu$ and the matrix $M_R$ does not break either $G_f$ or CP.  

For demonstration, we will consider a particular case with generator of $Z_2$ symmetry $Z \, = \, c^{n/2}$, where $c$ is one of the generators of the group $\Delta(6n^2)$ and the corresponding CP transformations reads $X(s)  = \ a \, b \, c^s \, d^{2 s} \, P_{23}$, where $s$ is a parameter that runs from $0$ to $(n-1)$. For this case, the form of $Y_D$ is given by 
\begin{equation}
\label{eq:YDgen}
Y_D \, = \, \Omega ({\bf 3}) \, R_{13} (\theta_L) \, \left(
\begin{array}{ccc}
 y_1 & 0 & 0\\
 0 & y_2 & 0\\
 0 & 0 & y_3
 \end{array}
 \right) \, R_{13} (-\theta_R) \, \Omega ({\bf 3^\prime})^\dagger \; .
\end{equation}
where angles $\theta_L$ and $\theta_R$ are free parameters,
with values in the range $[0,\pi)$. $\Omega(s) ({\bf 3})$ is a unitary matrix derived from $X ({\bf 3}) (s)$  
\begin{equation}
\label{case1Omegain3}
\Omega(s) ({\bf 3}) \, = \, e^{i \, \phi_s} \, U_{\mathrm{TB}} \,
\left( \begin{array}{ccc}
 1 & 0 & 0 \\
 0 & e^{-3 \, i \, \phi_s} & 0\\
 0 & 0 & -1
\end{array}
\right) \, ,
\end{equation}
with $\phi_s=\pi s/n$ and the TB form for~\cite{Harrison:2002er} 
\begin{equation}
U_{\mathrm{TB}} \, = \, 
 \left( \begin{array}{ccc}
  \sqrt{2/3} & \sqrt{1/3} & 0\\
 -\sqrt{1/6} & \sqrt{1/3} & \sqrt{1/2} \\
 -\sqrt{1/6} & \sqrt{1/3} & -\sqrt{1/2} 
 \end{array}
 \right) \, .
\end{equation}
Finally, we can find form for the PMNS mixing matrix
\begin{align}
    U \, = \, \Omega(s) ({\bf 3}) \, R_{13}(\theta_{\rm bf})\, K_\nu \, ,
    \label{eq:PMNS1}
\end{align}
where $\theta_{\rm bf}$ is a free real parameter (related to $\theta_L$) which is chosen to reproduce the best-fit values of the measured lepton mixing angles. $K_\nu$ is a diagonal matrix with entries equal to $\pm 1$ and $\pm {i}$, and ensures that the neutrino masses are non-negative.

\subsubsection{Production of RHNs}
RHNs can be as light as the electroweak scale in the resonant leptogenesis mechanism which makes it  testable in current and near-future experiments~\cite{Chun:2017spz}.
In the minimal type-I seesaw for our TeV-scale RHNs scenario, light neutrino masses and mixing constrain the Yukawa couplings to be highly suppressed $y_f\lesssim 10^{-7}$. This in turn suppresses the RHN production cross section at LHC ~\cite{CMS:2018jxx, ATLAS:2019kpx} for the smoking-gun signal of same-sign dilepton plus two jets without missing transverse energy~\cite{Keung:1983uu, Datta:1993nm, Han:2006ip, delAguila:2007qnc,Atre:2009rg, Dev:2013wba, Alva:2014gxa, Das:2015toa, Gluza:2015goa, Das:2016hof,Gluza:2016qqv,Das:2017gke,}.

\begin{figure}[t!]
\centering
\includegraphics[width=0.4\textwidth]{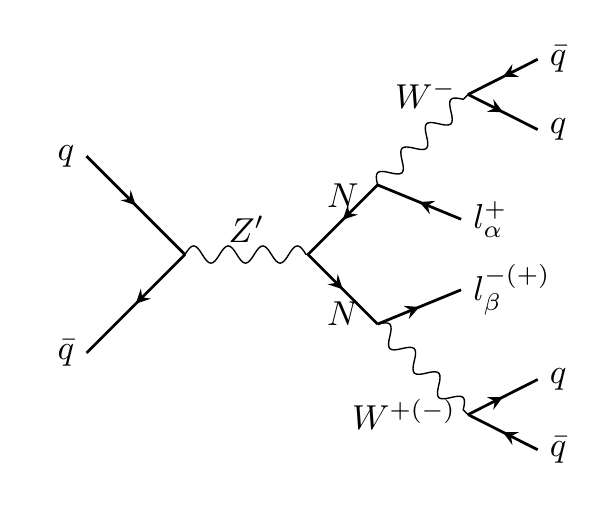}
\caption{Lepton Number Violating (LNV) signature of RHNs at hadron colliders in the $U(1)_{B-L}$ model. The final-state lepton flavors are dictated by the Yukawa coupling structure $Y_D$, which is governed by the choice of the flavor group and the CP transformation. Figure adapted from \cite{Chauhan:2021xus} under CC BY 4.0 license.}
\label{fig:feyn} 
\end{figure}

Therefore, one way for a more efficient production mechanism for the collider tests of the RHNs is to extend the SM gauge group, and make the RHNs as well as the SM quarks (and leptons) charged under this new group, leading to RHN production via the new gauge bosons~\cite{Deppisch:2015qwa}.  Consider for example, simple $U(1)_{B-L}$ extension of the SM~\cite{Davidson:1978pm, Marshak:1979fm} in which the RHNs can be pair-produced via the $U(1)_{ B-L}$ $Z'$ mediated process: $pp\to Z'\to N_i N_i$~\cite{Buchmuller:1991ce,Basso:2008iv, FileviezPerez:2009hdc, Kang:2015uoc, Cox:2017eme, Han:2021pun} (see Figure~\ref{fig:feyn}).
 
\begin{figure}[t!]
\centering
\includegraphics[width=0.6\textwidth]{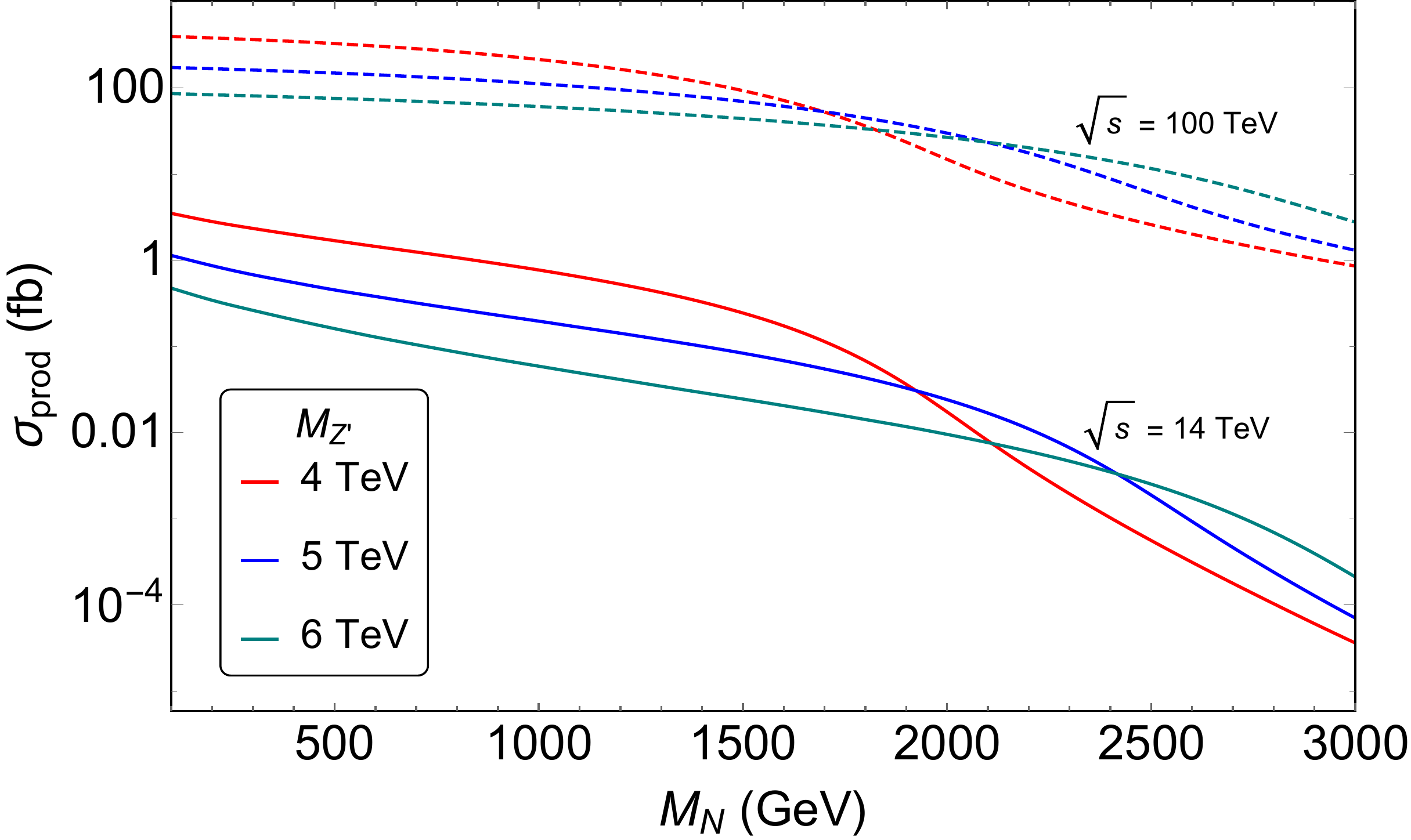}
\caption{ Normalized production cross-sections (setting $g_{B-L}=1$) for $pp\to Z' \to N_iN_i$, as a function of the RHN mass $M_N$ at $\sqrt s=14$ TeV LHC (solid lines) and $\sqrt s=100$ TeV FCC-hh (dotted lines) for $M_{Z'}= 4$ TeV (red), 5 TeV (blue) and 6 TeV (green). Figure adapted from \cite{Chauhan:2021xus} under CC BY 4.0 license.}
\label{fig:collider}
\end{figure}
We calculate the hadron collider production cross sections and results for the gauge coupling $g_{B-L}=1$ are shown in Figure~\ref{fig:collider} for three different values of $M_{Z'}=$ 4,5,6 TeV (red, blue and green, respectively). The solid lines are for the LHC center-of-energy $\sqrt s=14$ TeV, whereas the dotted lines are for a future 100 TeV collider. We find that it is challenging to find significant number of events even at the HL-LHC, due to the cross section being below fb-level. Whereas the 100 TeV collider option will have far better sensitivity and sizable number of events.

\subsubsection{Decay Lengths}
\label{subsec:RHnuwidths}
\begin{figure}[t!]
\centering
\includegraphics[width=0.49\textwidth]{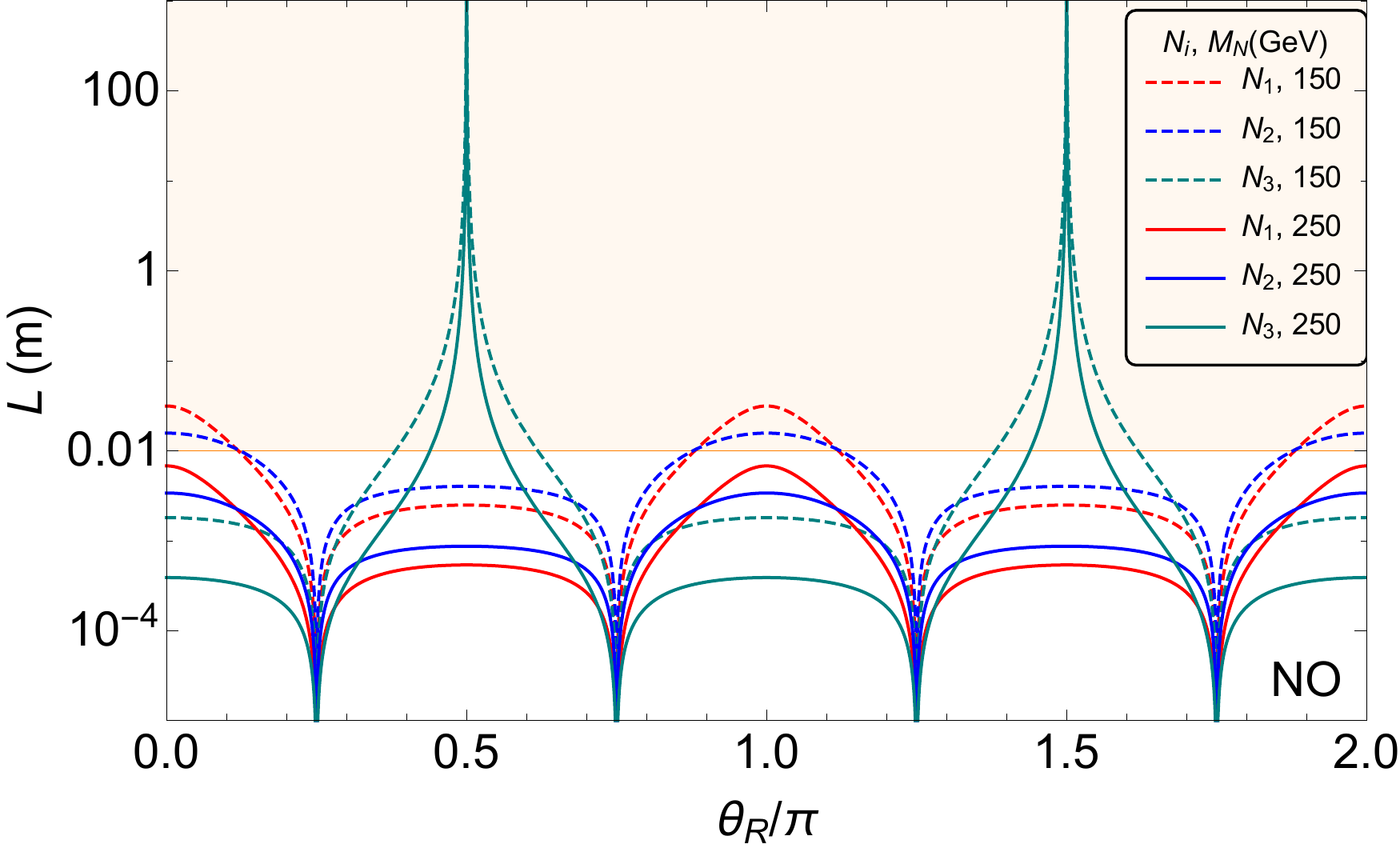}
\includegraphics[width=0.49\textwidth]{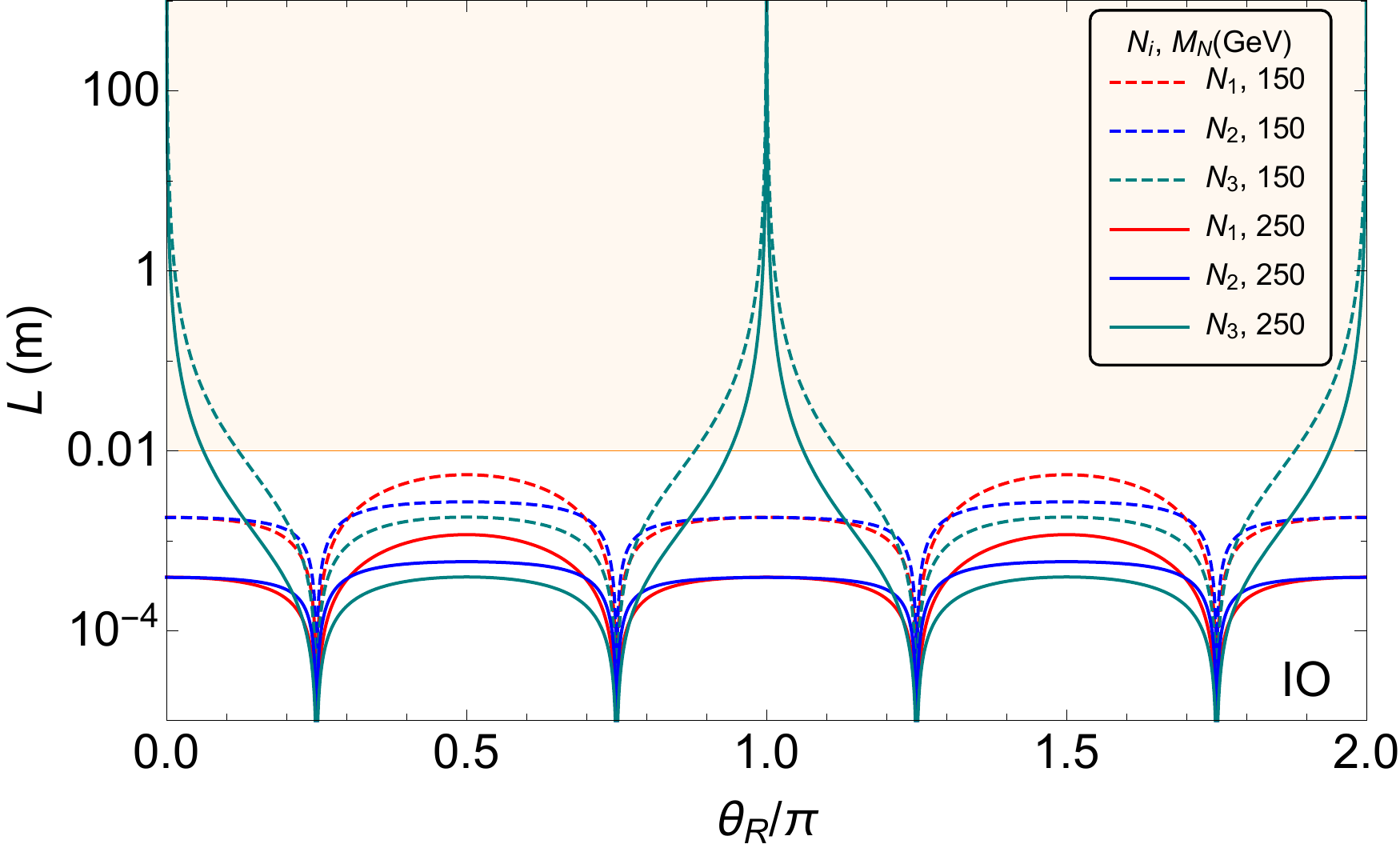}\\
\caption{ Decay lengths for $N_{1,2,3}$ are plotted against $\theta_R$ for different values of the RHN mass scale $M_N$ ( with $m_0=0$) The left (right) panels are for NH (IH). The shaded (unshaded) region indicates the displaced/long-lived (prompt) signal regime. Figure adapted from \cite{Chauhan:2021xus} under CC BY 4.0 license.}
\label{fig:dln1}
\end{figure}

After being produced through the $Z'$ mediated process, the RHNs decay into SM final states. Since the total decay width $ \Gamma_i$ of the RHN $N_i$ at tree-level depend on $Y_D$, given by
\begin{equation}
    \Gamma_i \, = \, \frac{(Y_D^\dagger \, Y_D)_{ii}}{8 \, \pi } M_i \, ,
\end{equation}
Thus the RHNs decay lengths would depend on the choice for generator $Z$ of the $Z_2$ symmetry and the choice of the CP transformation $X$. For the case in consideration, the expressions for the decay widths are independent of values of $s$, and only depend on the Yukawa couplings $y_f$ and the angle $\theta_R$: 
\begin{subequations}\label{gammaC1}
\begin{alignat}{2}
\Gamma_1 \, & =  \, \frac{M_N}{24 \, \pi} \, \left( 2\, y_1^2 \, \cos^2 \theta_R + y_2^2 + 2\, y_3^2 \, \sin^2 \theta_R \right)  \, , \\
\Gamma_2 \, & =  \,  \frac{M_N}{24 \, \pi} \, \left( y_1^2 \, \cos^2\theta_R + 2 \, y_2^2 + y_3^2\, \sin^2 \theta_R  \right) \, ,  \\
\Gamma_3 \ & =  \, \frac{M_N}{8 \, \pi} \, \left( y_1^2 \, \sin^2 \theta_R + y_3^2 \, \cos^2 \theta_R \right) \, .
\end{alignat}
\end{subequations}
The boost factor for decay lengths in the laboratory frame is obtained by assuming that the RHNs are produced via $Z^\prime$ with  mass $M_{Z^\prime}=4$ TeV, and plotted as a function of $\theta_R$ in Figure~\ref{fig:dln1}.
 
As can be seen in the expressions, for strong NH and strong IH corresponding to $y_1=0$ and $y_3=0$ respectively (when i.e.~$m_0=0$), the ERS points for NH are $\theta_R \to \pi/2, \, 3 \pi/2$ and  $\theta_R \to 0, \, \pi$ for IH. At ERS points, $N_3$ becomes long-lived as we see from Eq.~\eqref{gammaC1} that $\Gamma_3\to 0$.  Given a sufficiently large production cross section, $N_3$ can be searched with displaced vertex signature at the LHC or with dedicated LLP detectors. Most signals from $N_{1,2}$ decays are prompt but can also be displaced depending on the choice of $\theta_R$.  

\subsubsection{Branching Ratios}

After the on-shell production of RHNs, underlying Yukawa structure predicts their decay BRs. Since BRs are dependent on $Y_D$, the choice for generator $Z$ of the $Z_2$ symmetry and the choice of the CP transformation $X$ directly governs the BRs. Assuming RHN are heavy enough, $N_i$ can decay into $\ell_\alpha\, W,\,\nu_\alpha\, Z$ and $\nu_\alpha\, H$, through its mixing with the SM neutrinos, with the partial decay widths proportional to $\left|\left(Y_D\right)_{\alpha i}\right|^2$.

Consider the decay of long-lived $N_3$ at a LLP detector with $m_0=0$ and $M_N=250$ GeV, we find the following ratio of the BRs
\begin{equation}
\label{eq:br3}
{\rm BR}(N_3\to e^\pm W^\mp): {\rm BR}(N_3\to \mu^\pm W^\mp): {\rm BR}(N_3\to \tau^\pm W^\mp) \, = \, \left\{\begin{array}{ll}
1: 27.7:18.1 & ({\rm NH}) \\
8.5:1:3.7 & ({\rm IH})
\end{array}\right. ,
\end{equation}
independent of $\theta_R$ and $s$, and almost independent of $M_N$, if $M_N \gg m_W$. 

A test of the neutrino mass hierarchy can be done by measuring these BRs for at least two charged lepton flavors $\alpha$ at an LLP detector. Decays signals of $N_{1,2}$ at LHC (prompt or displaced vertex) can also be used to test mass hierarchy but specifically depend on $\theta_R$ as well as on the chosen CP symmetry $X (s)$.
 
\subsubsection{Same-sign dilepton signals}
\label{subsec:SSleptonsLHCRHnu}
As shown in Figure~\ref{fig:feyn}, once produced via the $Z'$, the decay of Majorana $N_i$'s into charged leptons leads to a Lepton Number Violating (LNV) as well as Lepton Flavor Violating (LFV) signal (only if $\alpha\neq \beta$) ~\cite{Deppisch:2013cya},
\begin{align}
    pp \to Z'\to N_iN_i\to \ell_\alpha^\pm \ell_\beta^\pm +2W^\mp \to \ell_\alpha^\pm \ell_\beta^\pm +4j \, .
    \label{eq:signal}
\end{align}
The above process~\eqref{eq:signal} has a much smaller SM background than its lepton number conserving counterpart. Since the partial decay widths of RHN dependent on $Y_D$, the LNV signal cross-section is affected by the choice for generator $Z$ of the $Z_2$ symmetry and the choice of the CP transformation $X$.

By constructing simple observables out of the same-sign dilepton charge asymmetry, we can probe the high-energy CP  phases in the Yukawa coupling matrix at colliders. 

In particular, we can define two observables $\sigma_{\rm LNV}^{\alpha,-}$ (difference) and  $\sigma_{\rm LNV}^{\alpha,+}$ (sum) of the same-sign charged-lepton final states of a given flavor $\alpha$ 
\begin{align}
    \sigma_{\rm LNV}^{\alpha,\pm} \, = \, & \sum_i \sigma_{\rm prod}(pp\to N_iN_i)\left(\left[{\rm BR}(N_i\to \ell_\alpha^-W^+)\right]^2\pm \left[{\rm BR}(N_i\to \ell_\alpha^+W^-)\right]^2 \right) \nonumber \\
    & \quad  \times \left[{\rm BR}(W\to jj)\right]^2 \, .
\end{align}
The flavored CP asymmetries $\varepsilon_{i\alpha}$ can then be related to the ratio $\sigma_{\rm LNV}^{\alpha,-}/\sigma_{\rm LNV}^{\alpha,+}$, See refs.~\cite{Bray:2007ru, Blanchet:2009bu, Dev:2019ljp}. Thus, measuring $\sigma_{\rm LNV}^{\alpha,-}/\sigma_{\rm LNV}^{\alpha,+}$ can help measure the CP asymmetry which is determined by the group theory parameters. 

\begin{figure}[t!]
\centering
\includegraphics[width=0.49\textwidth]{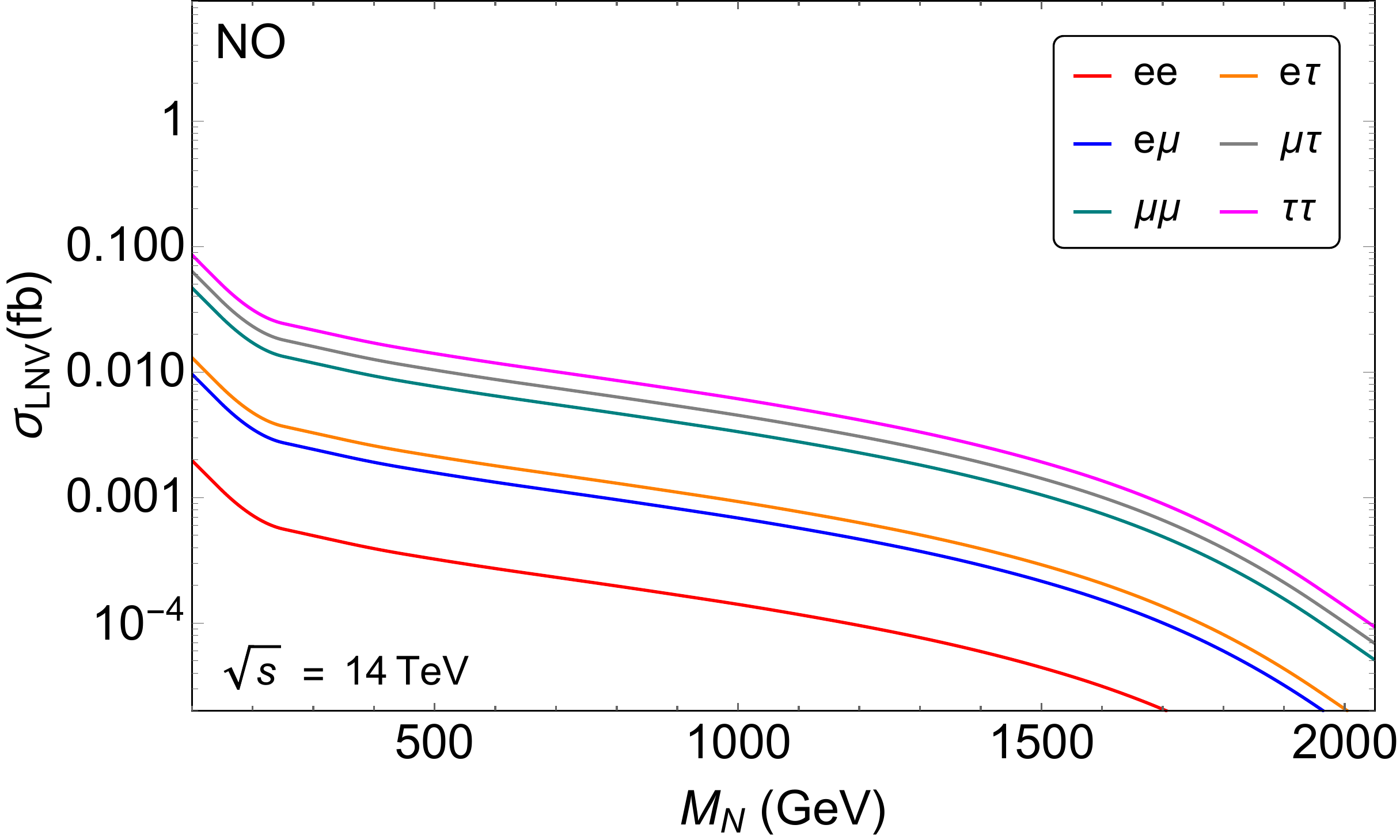}
\includegraphics[width=0.49\textwidth]{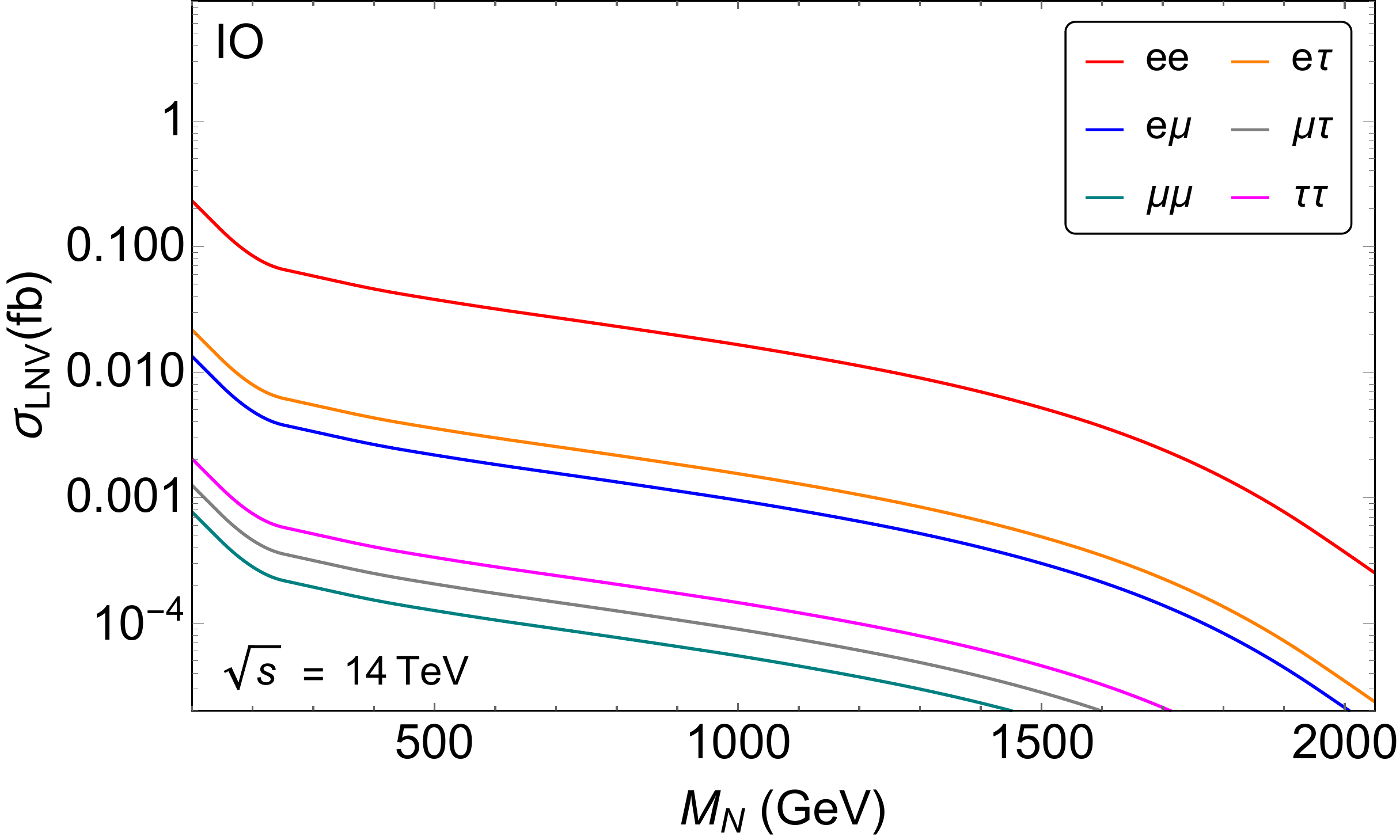}
\caption{Normalized LNV signal ($g_{B-L}=1$) as a function of the RHN mass scale $M_N$ at $\sqrt s=14$ TeV LHC for all possible lepton flavor combinations in the strong NH (left) and strong IH (right) limit for $M_{Z'}= 4$ TeV. Figure adapted from \cite{Chauhan:2021xus} under CC BY 4.0 license.}
\label{fig:collider1}
\end{figure}

\subsubsection{Collider prospects and leptogenesis}
\label{subsec:colliderleptogenesis}

We can now compute the baryon asymmetry $\eta_B$ in our scenario by first evaluating the CP asymmetry, given by 
\begin{align}
    \varepsilon_i \ \equiv \ \sum_\alpha \varepsilon_{i\alpha} \, = \, \frac{1}{8\pi \left(Y_D^\dag Y_D\right)_{ii}} \sum_{j\neq i} {\rm Im}\left[\left(Y_D^\dag Y_D\right)_{ij}\right] {\rm Re}\left[\left(Y_D^\dag Y_D\right)_{ij}\right]{\cal F}_{ij}  \, , 
\label{eq:eps3}
\end{align}
where
\begin{align}
    {\cal F}_{ij} \, = \, \frac{M_i M_j (M_i^2-M_j^2)}{(M_i^2-M_j^2)^2+A_{ij}^2} \, 
\end{align}
Since the CP asymmetry is a function of $Y_D$, the choice for generator $Z$ of the $Z_2$ symmetry and the choice of the CP transformation $X$ determines the form of $\varepsilon_i$ and in turn the $\eta_B$.

\begin{figure}[t!]
\centering
\includegraphics[scale=0.45]{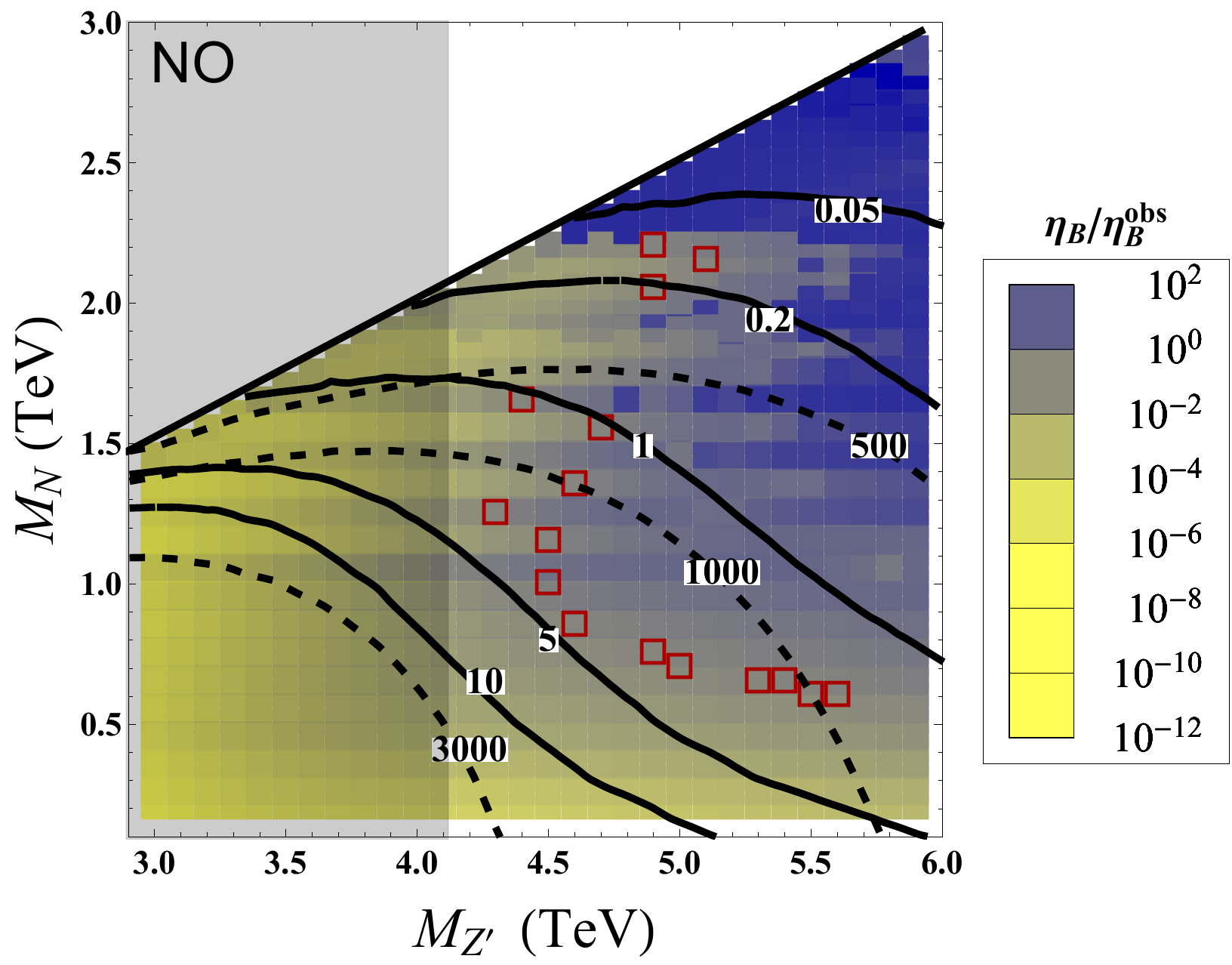}
\includegraphics[scale=0.45]{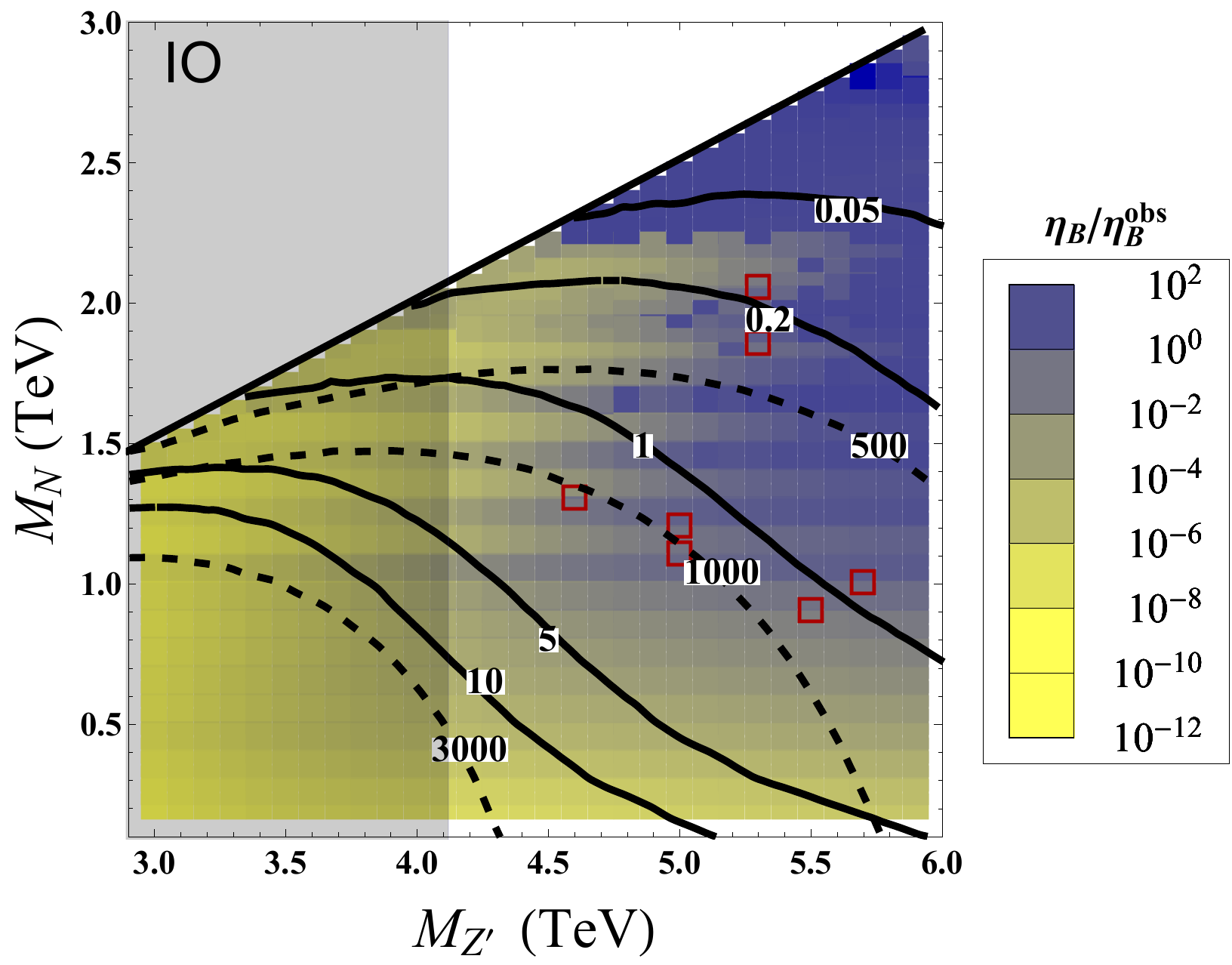}
\caption{Baryon asymmetry $\eta_B$ predicted relative to the observed value $\eta_B^{\rm obs}$ in the $(M_{Z'},M_N)$ plane for a fixed $g_{B-L}=0.1$, $n=26$ and $s$ set to $2 \ (17)$ for strong NH (IH) in the left (right) panel, with $\theta_R$ set to respective ERS points. The red boxes correspond to $\eta_B$ within $10\%$ of $\eta_B^{\rm obs}$. The contours show $\sigma_{\mathrm{prod}}$ (in ab) at the $\sqrt s=14$ TeV LHC (solid) and at $\sqrt s=100$ TeV FCC-hh (dashed). Figure adapted from \cite{Chauhan:2021xus} under CC BY 4.0 license.}
\label{fig:etaBZp}
\end{figure}  

Figure~\ref{fig:etaBZp} shows the contours of $\sigma(pp\to Z'\to N_iN_i)$ in ab for both $\sqrt s=14$ TeV LHC (solid lines) and $\sqrt s=100$ TeV FCC-hh (dashed lines). We discuss the results for $\eta_B$ from resonant leptogenesis with the collider cross-sections in the $(M_{Z'},M_N)$ plane. Successful leptogenesis for strong NH yields  $\sigma_{\mathrm{prod}}\lesssim 5$ ab at $\sqrt s=14$ TeV LHC. Even after assuming a fairly low SM background and the decay BRs, the number of events is still $<{\cal O}(1)$ with the final target luminosity of 3 ab$^{-1}$ at HL-LHC. The situation worsens for strong IH, where the cross sections are smaller by at least an order of magnitude compared to the strong NH case. For the region of successful leptogenesis, a future $100$ TeV collider can reach a $\sigma_{\mathrm{prod}}$ up to $2000$ ab. After assuming a fairly low SM background and taking the decay BRs into account with 30 ab$^{-1}$ integrated luminosity, we expect up to about 1000 LNV events. By going to higher $Z'$ masses, the detection prospects at 100 TeV collider can be improved significantly. This is due to relaxed experimental limits on $g_{B-L}$ for $M_{Z'}\gtrsim 6$ TeV~\cite{Das:2021esm}.
For instance at $M_{Z'}=7$ TeV, $g_{B-L}$ is allowed to be as large as one. Since $\sigma_{\mathrm{prod}}$ scales as $g_{B-L}^2$ for $M_N<M_{Z'}/2$, apart from a mild suppression due to change in the $Z'$ mass, the cross-section gets enhanced by a factor of 100. 

\subsection{Flavor symmetry and collider physics with $A_4$, $S_4$, $Q_6 \times Z_2$, $T_7$, $\ldots$}
\label{sec:otherpheno}

As another example, in \cite{CarcamoHernandez:2013yiy} another model has been considered  with two Higgs doublets  and flavor symmetry $A_{4} \times {Z}_2 \times { Z}^{\prime}_{2}$. In \cite{deMedeirosVarzielas:2015ybd} some phenomenological consequences of the model for collider physics and dark matter problem were further explored. 
The ${\mathbb Z}_2$-even scalar fields of the considered model gives two generations of fields that couple completely off-diagonally to the charged leptons thus the model predicts LFV processes $\tau \to 2 \mu e$, $\mu \to e \gamma$ and $e^+e^- \to \tau^+ \mu^-$. As a consequence, LFV processes can be also searched for at hadron colliders through the process $pp \to jj H_2 H_2 \to 2j4l$, $H_2$ is a non-standard Higgs scalar. The non-standard sector of the model is rich enough to include the Dark Matter fields candidates (scalars and Majorana neutrinos). As shown in the analysis, they can be probed at hadron colliders. In this way, it has been found that it is possible to account for the observed relic density for DM with a mass in the interval between 47 and 74 GeV or in the interval 600 GeV and 3.6 TeV.

From the present model and the model considered in section \ref{sec:delta6pheno}, it is clear that the analysis of the flavor symmetry leads to the rich phenomenology connected with intensity, energy frontiers as well as cosmological studies. A short note given in sections \ref{sec:delta6pheno} and in this section 
 on chosen groups does not exhaust the whole range of possibilities of phenomenological studies of discrete flavor symmetries and can be extended to another groups including, for instance, collider signatures of vector-like fermions with $Q_6 \times Z_2$ symmetry~\cite{Bonilla:2021ize} or
collider signature of $T_7$ flavor symmetry with gauged $U(1)_{B-L}$ \cite{Cao:2010mp} (so a mixture of discrete and continuous symmetries is also possible) or studies on LFV Higgs decays in context of 3HDM with $S_4$ symmetry \cite{Campos:2014zaa}.  In an unified framework~\cite{Belyaev:2018vkl}, based on  GUT $SU(5)\times A_4$ symmetry, studies on muon anomalous magnetic moment
$g-2$,  dark matter have been performed in the context of  LHC data where the right-handed smuon with masses are predicted to be around 100 GeV.

\section{New ideas}
 The studies of discrete family symmetries continue, with new ideas concerning pure theoretical aspects of models building as well as phenomenological studies. We comment here on some of the latest interesting concepts. 
 
 \subsection{ Flavor Symmetry and Gravitational Waves}
 
Gravitational waves observations provides us  a new window into the early universe. As the Majorana mass term can be generated by $B-L$ breaking at a high energy scale (GUT or beyound),  the related phase transition in the early universe can produce a possibly observable stochastic
gravitational wave background \cite{AlvesBatista:2021gzc}.
Thus, the emerging field of gravitational wave astronomy may path the way towards hints and the search for the origin of a Majorana neutrino mass term as well as flavor symmetries in the lepton sector. For recent studies on this subject we refer the reader to \cite{Gelmini:2020bqg} and references therein. In~\cite{Gelmini:2020bqg},  it was pointed out that breaking of non-Abelian discrete symmetries (here $A_4$)   may  provide a characteristic signature of lepton flavor models  at gravitational wave detectors such as ET, LIGO, DECIGO, BBO, LIGO, etc.  Such studies could provide one of the  important probes to  test the flavor models.

\subsection{Flavor symmetry and neutrino experiments}

As discussed earlier there exists a wide class of models with various discrete flavor symmetry groups $G_f$. Now, to probe such symmetry, one needs to find the possibilities of testing its predictions at the current and future neutrino experiments. Such studies crucially depend on the breaking pattern of $G_f$ into its residual subgroups for charged lepton sector $G_e$ and neutrino sector $G_\nu$.  For example, in~\cite{Blennow:2020snb}, the authors have studied implication of breaking of $G_f \times CP$ (with $G_f=A_4, S_4, A_5$) into $G_e > Z_2$, $G_{\nu}=Z_2 \times CP$ in the context of ESSnuSB experiment~\cite{ESSnuSB:2013dql}. Such breaking pattern is usually observed in the semi-direct approach of flavor model building~\cite{King:2019gif}.
In a similar approach~\cite{Blennow:2020ncm,Petcov:2018snn}, with considered breaking pattern e.g.  $G_e = Z_k, k > 2$ or $Z_m \times Z_n$, $m, n \geq 2$
and  $G_{\nu} = Z_2 \times CP$ residual symmetry for charged lepton and neutrino sector respectively in the context of ESSnuSB, T2HK, DUNE, and JUNO experiments. In each cases distinct  constraints were obtained on the neutrino oscillation parameters $\delta_{CP},\theta_{12}$ and  $\theta_{23}$.
In another approach~\cite{Wang:2018dwk}, the authors have demonstrated an approach to construct operators for neutrino non-standard interactions based on $A_4$ discrete symmetry and study its feasibility at the DUNE experiment. Furthermore, for  studies of consequences of partial $\mu-\tau$ reflection symmetry at DUNE and Hyper-Kamiokande see ~\cite{Chakraborty:2018dew,Chakraborty:2019rjc}. Guided by the considered discrete flavor symmetry $G_f$, sum rules involving neutrino masses and mixing may also have inherent characteristics~\cite{Gehrlein:2016wlc, Gehrlein:2016fms,Gehrlein:2017ryu} to be confronted with the neutrino experiments mentioned here.  

\subsection{Modular Symmetry}
 
\noindent

Recently, in an appealing proposal to understand the flavor pattern of the fermions, the idea of modular symmetry~\cite{Altarelli:2005yp,Feruglio:2017spp} has been reintroduced.
There exist a plethora of models based on various non-Abelian discrete flavor symmetries and other various finite groups. The spectrum of the models here sometime is so large that it is difficult to obtain a clear clue of the underlying flavor symmetry. Additionally, there are a few major disadvantages of using this conventional approach. Firstly, the effective Lagrangian of a typical flavor model is given by introducing a large set of flavons. Secondly, typical vacuum alignment of these flavons is required which essentially determine the flavor structure of quark, leptons and certain sometimes auxiliary symmetries are also needed to forbid unwanted operators contributing in the mass matrix.  The third and most crucial disadvantage of the use of the conventional approach is the breaking sector of flavor symmetry typically introduces many unknown parameters hence compromising the minimality. On the contrary, the primary advantage of models with modular symmetry~~\cite{Altarelli:2005yp,Feruglio:2017spp} is that the flavon fields might not be needed and the flavor symmetry can be uniquely broken by the vacuum expectation value of the modulus $\tau$.  Here the Yukawa couplings are written as modular forms, functions of only one complex parameter i.e., the modulus $\tau$, and transform non-trivially under the modular symmetry. Furthermore, all the higher-dimensional operators in the superpotential are completely determined by modular invariance if supersymmetry is exact hence auxiliary Abelian symmetries are not needed in this case. Just like the conventional model building approach models with modular symmetry can also be highly predictive. The fundamental advantage being the neutrino masses and mixing parameters can be predicted in terms of a few input parameters. There are already several activities adopting this approach, for a few examples see~\cite{Kobayashi:2018vbk,Penedo:2018nmg,Kobayashi:2018wkl,deAnda:2018ecu,Okada:2018yrn,Novichkov:2018ovf,Ding:2019xna,Wang:2020dbp,Wang:2019ovr,Wang:2019xbo,Wang:2020lxk,Wang:2021mkw}. As many models make it possible to understand the fermionic  
mixing, still there is a scope for the study of its phenomenological aspects such as its possible connection with matter-antimatter asymmetry and dark matter, among others. However, testing the modular symmetry in the context of current and future current neutrino experiments like T2K, NO$\nu$A and DUNE is very hard as obtain  robust correlations among the neutrino oscillation parameters. However, sum rules obtained in
models based on modular symmetries with residual symmetries may shade some line in this regard~\cite{Gehrlein:2020jnr}. 
 
\subsection{Inverse matrix problem and family symmetries \label{sec:invprob}}
An inverse eigenvalue (singular value) problem is a method which recreates a matrix from a given spectrum \cite{Chu_1998,chu_golub_2002}. This approach can be used for modeling neutrino mass and mixing matrices. It can be applied in studies of discrete symmetries when one starts from constructing a given mass or mixing matrix, whose spectrum agrees with the experiment, and then tries to find a corresponding symmetry associated to the obtained structure. It is important to notice that the structure of matrices obtained in this way is not arbitrary but is connected with majorization relations between matrix elements and spectrum. 
In that way the possible structures are restricted, e.g. to ensure the positive heavy masses of sterile neutrinos in CP invariant seesaw models the diagonal elements of the $M_{R}$ submatrix must be much larger than off-diagonal elements, see Theorem III.1 and discussion in \cite{Flieger:2020lbg}.  After the recent rediscovery in context of neutrino physics of a simple relation between eigenvalues and eigenvectors \cite{Denton:2019ovn, Denton:2019pka}
\begin{equation}
\vert v_{ij} \vert^{2} \prod_{k=1, k\neq i}^{n}( \lambda_{i}(A) - \lambda_{k}(A)) = 
\prod_{k=1}^{n-1}( \lambda_{i}(A) - \lambda_{k}(M_{j}))
\end{equation}
where $v_{ij}$ is the $j^{th}$ component of the eigenvector $v_{i}$, $\lambda_{i}(A)$ $i=1,...,n$ are increasingly ordered eigenvalues of the $n\times n$ Hermitian matrix $A$ and $\lambda_{k}(M_{j})$ $k=1,...,n-1$ are eigenvalues of the submatrix $M_{j}$ obtained by deleting $j^{th}$ row and column from $A$, the inverse problem methods allow for simultaneous modeling of mass and mixing matrices. This observation can lead to further studies of the neutrino discrete family symmetries both in the light and heavy neutrino sectors with or without heavy neutrino mass degeneracy.

\section{Summary and Outlook}

In the paper, we consider neutrino mass, mixing and flavor model
building strategies based on discrete family symmetries.
We consider fixed patterns (BM, TBM, GR, HG) and more elaborated symmetry groups with unbroken residual symmetries which are a consequence of recent experimental results where nonzero neutrino mixing angle $\theta_{13}$ has been determined and the Dirac CP phase is preferable non-zero for both NH and IH mass scenarios.  Examples of such groups are among others $A_4,  S_4, T'$,  $\Delta(27)$ and $ A_5$. We discuss also flavor symmetry and mechanisms of mass generation, and flavor symmetry in multi-Higgs doublet models. 

As there are plenty of possible flavor symmetry models, a natural question is how to falsify or validate these models?  Generally, such symmetries are broken at a high scale and beyond our experimental reach. Nonetheless, phenomenological study connected with flavor symmetry effects is rich and various models can be probed effectively in various low energy intensity frontier or high energy collider frontiers. One of the key tests for such models comes from RHNs. Their Yukawa couplings are relevant for collider signals, BAU, and leptogenesis. We give an example of such studies for the $\Delta(6n^2)$ group, other groups are also discussed in the context of LFV and DM effects. 

The concept of flavor symmetry is developing all the time. In this respect, we should mention possible connections with gravitational waves. New concepts related to family symmetries come also from modular symmetries or the inverse eigenvalue problem. With neutrino physics entering the precision era, there is an exciting prospect for more intensive studies of flavor symmetries in the future both for Majorana and Dirac neutrino scenarios.

\acknowledgments
This work has been supported in part by the Polish National Science Center (NCN) under grant 2020/37/B/ST2/02371 and the Research Excellence Initiative of the University of Silesia in Katowice. 
 
\bibliographystyle{elsarticle-num-ID}
\bibliography{ref,refs,2loops,refGC}
\end{document}